\documentclass[preprint,12pt,3p]{elsarticle}

\usepackage{graphics}
 \usepackage{graphicx}
 \usepackage{subfig}
\usepackage{amssymb}
\usepackage{amsfonts}
\usepackage{amsmath}
\usepackage{color}
\usepackage{amssymb}

\def\spacce#1{\hskip #1pt}
\def\drawline#1#2{\raise 2.5pt\vbox{\hrule width #1pt height #2pt}}

\def\solid{\drawline{22}{1.0}\nobreak\ }

\def\tdash{\hbox{\drawline{4}{1.0}\spacce{2}}}
\def\dashed{\tdash \tdash \tdash \tdash \nobreak\ }

\def\tdot{\hbox{\drawline{1}{1.0}\spacce{2}}}
\def\dotted{\tdot \tdot \tdot \tdot \tdot \tdot \tdot \tdot \nobreak\ }

\def\dashdot{\tdash \tdot \tdash \tdot \tdash \nobreak\ }

\def\ssquare{${\vcenter{\hrule height 0.9pt
       \hbox{\vrule width 0.9pt height 10pt \kern 10pt
       \vrule width 0.9pt}
       \hrule height 0.9pt}}$\nobreak\ }
       
\def\ssquareb{$\Box$\nobreak\ }       
       
\def\blackssquare{$\scriptstyle\blacksquare$\nobreak\ }
       
\def\circle{$\circ$\nobreak\ }

\def\blackcircle{$\bullet$\nobreak\ }

\def\losange{$\Diamond$\nobreak\ }

\def\blacklosange{$\blacklozenge$\nobreak\ }

\def\trianup{\raise 1.25pt\hbox{$\triangle$}\nobreak\ }

\def\blacktrianup{\raise 1.25pt\hbox{$\blacktriangle$}\nobreak\ }

\def\bigtriandown{\raise 1.25pt\hbox{$\bigtriangledown$}\nobreak\ } 

\def\plus{$+$ \nobreak}

\def\asterix{$\ast$ \nobreak}

\def\solidopencircle{\solid \nobreak\spacce{-19.5}\raise
 -0.5pt\hbox{\circle} \nobreak\ }
 
\def\solidblackcircle{\solid \nobreak\spacce{-21}\raise
 -0pt\hbox{\blackcircle} \nobreak\ }

\def\solidopensquare{\solid \nobreak\spacce{-19}\raise
 -1.0pt\hbox{\ssquare} \nobreak\ }
 
\def\solidblacksquare{\solid \nobreak\spacce{-19}\raise
 -1.0pt\hbox{\blackssquare} \nobreak\ }
 
\def\solidopenlosange{\solid \nobreak\spacce{-19}\raise
 -1.0pt\hbox{\losange} \nobreak\ }
 
\def\solidblacklosange{\solid \nobreak\spacce{-21}\raise
 -2.0pt\hbox{\blacklosange} \nobreak\ }

\def\solidopentriangleup{\solid \nobreak\spacce{-19}\raise
 -1.pt\hbox{\trianup} \nobreak\ }
 
\def\solidblacktriangleup{\solid \nobreak\spacce{-21.5}\raise
 -2.0pt\hbox{\blacktrianup} \nobreak\ }

\def\solidopencross{\solid \nobreak\spacce{-22}\raise
 -1.0pt\hbox{\cross} \nobreak\ }
 
\def\solidopenplus{\solid \nobreak\spacce{-21}\raise
 -1.5pt\hbox{\plus} \nobreak\ }
 
\def\solidopenasterix{\solid \nobreak\spacce{-21}\raise
 -1.5pt\hbox{\asterix} \nobreak\ }

\def\dashedopencircle{\dashed \nobreak\spacce{-22}\raise
 -1.0pt\hbox{\circle} \nobreak\ }
 
\def\dashedblackcircle{\dashed \nobreak\spacce{-23}\raise
 -1.4pt\hbox{\blackcircle} \nobreak\ }
 
\def\dashedopensquare{\dashed \nobreak\spacce{-22}\raise
 -1.0pt\hbox{\ssquare} \nobreak\ }
 
\def\dashedblacksquare{\dashed \nobreak\spacce{-23}\raise
 -1.0pt\hbox{\blackssquare} \nobreak\ }
 
\def\dashedopenlosange{\dashed \nobreak\spacce{-22}\raise
 -1.0pt\hbox{\losange} \nobreak\ }
 
\def\dashedblacklosange{\dashed \nobreak\spacce{-23}\raise
 -2.0pt\hbox{\blacklosange} \nobreak\ }
 
\def\dashedopentriangleup{\dashed \nobreak\spacce{-22}\raise
 -1.0pt\hbox{\trianup} \nobreak\ }
 
\def\dashedblacktriangleup{\dashed \nobreak\spacce{-24}\raise
 -2.0pt\hbox{\blacktrianup} \nobreak\ }
 
\def\dashedopencross{\dashed \nobreak\spacce{-24}\raise
 -1.0pt\hbox{\cross} \nobreak\ }
 
\def\dashedopenplus{\dashed \nobreak\spacce{-24}\raise
 -1.0pt\hbox{\plus} \nobreak\ }

\def\dashedopenasterix{\dashed \nobreak\spacce{-22}\raise
 -1.5pt\hbox{\asterix} \nobreak\ }

\def\dottedopencircle{\dotted \nobreak\spacce{-22}\raise
 -1.0pt\hbox{\circle} \nobreak\ }
 
\def\dottedblackcircle{\dotted \nobreak\spacce{-23}\raise
 -1.4pt\hbox{\blackcircle} \nobreak\ }

\def\dottedopensquare{\dotted \nobreak\spacce{-22}\raise
 -1.0pt\hbox{\ssquare} \nobreak\ }
 
\def\dotted\mathbfuare{\dotted \nobreak\spacce{-23}\raise
 -1.0pt\hbox{\blackssquare} \nobreak\ }
 
\def\dottedopenlosange{\dotted \nobreak\spacce{-22}\raise
 -1.0pt\hbox{\losange} \nobreak\ }
 
\def\dottedblacklosange{\dotted \nobreak\spacce{-24}\raise
 -2.0pt\hbox{\blacklosange} \nobreak\ }
 
\def\dottedopentriangleup{\dotted \nobreak\spacce{-22}\raise
 -1.0pt\hbox{\trianup} \nobreak\ }
 
\def\dottedblacktriangleup{\dotted \nobreak\spacce{-25}\raise
 -2.0pt\hbox{\blacktrianup} \nobreak\ }
 
\def\dottedopencross{\dotted \nobreak\spacce{-23.5}\raise
 -1.0pt\hbox{\cross} \nobreak\ }
 
\def\dottedopenplus{\dotted \nobreak\spacce{-23.5}\raise
 -1.0pt\hbox{\plus} \nobreak\ }

\def\dottedopenasterix{\dotted \nobreak\spacce{-23.5}\raise
 -1.0pt\hbox{\asterix} \nobreak\ }

\def\whitehisto{${\vcenter{\color{black} \hrule height 1.0pt
       \hbox{\vrule width 1.0pt height 4pt \kern 8pt
       \vrule width 1.0pt}
       \hrule height 1.0pt}}$\nobreak\ }
       
\def\histosymb#1{${\vcenter{\color{#1} \hrule height 0.0pt
       \hbox{\vrule width 11.0pt height 6pt \kern 0pt 
       \vrule width 0.0pt}
       \hrule height 0.0pt}}$\nobreak\ }

\def\symbol#1#2#3#4#5#6{\color{#5}#1 \nobreak\spacce{-#3}\raise
 -#4pt\hbox{\color{#6}#2}\color{black} \nobreak\ }

\usepackage{calrsfs}
\newcommand*\colvec[3][]{    \begin{pmatrix}\ifx\relax#1\relax\else#1\\\fi#2\\#3\end{pmatrix}}
\newcommand{\R}{\mathbb{R}}

\journal{Journal Computational Physics}

\begin{document}

\begin{frontmatter}

\title{A general formulation of Bead Models applied to flexible fibers and active filaments at low Reynolds number}

\author[label1,label2]{Blaise Delmotte}
\address[label1]{University of Toulouse – INPT-UPS: Institut de M\'ecanique des Fluides, Toulouse, France.}
\address[label2]{IMFT - CNRS, UMR 5502 1, All\'ee du Professeur Camille Soula, 31 400 Toulouse, France.}

\cortext[cor1]{Corresponding author: Franck Plourabou\'e. Tel.: +33 5 34 32 28 80}

\ead{blaise.delmotte@imft.fr}

\author[label1,label2]{Eric Climent}
\ead{eric.climent@imft.fr}

\author[label1,label2]{Franck Plourabou\'e \corref{cor1}}
\ead{franck.plouraboue@imft.fr}

\begin{abstract}
This contribution provides a general framework to use Lagrange multipliers for the simulation of low Reynolds number fiber dynamics based on Bead Models (BM). This formalism provides an efficient method to account for kinematic constraints. We illustrate, with several examples, to which extent the proposed formulation offers a flexible and versatile framework for the quantitative modeling of flexible fibers deformation and rotation in shear flow, the dynamics of actuated filaments and the propulsion of active swimmers.  Furthermore, a new contact model called Gears Model is proposed and successfully tested. It avoids the use of numerical artifices such as repulsive forces between adjacent beads, a source of numerical difficulties in the temporal integration of previous Bead Models.

\end{abstract}


\end{frontmatter}

% \linenumbers

\section{Introduction}
\label{}
 The dynamics of solid-liquid suspensions is a longstanding topic of research while it combines
difficulties arising from the coupling of multi-body interactions in a viscous fluid with possible
deformations of flexible objects such as fibers. A vast literature exists on the response of suspensions of
solid spherical or non-spherical particles due to its ubiquitous interest in natural and industrial 
processes. When the objects have the ability to deform many complications arise. The coupling
between suspended particles will depend on the positions (possibly orientations) but also on the shape of 
individuals, introducing intricate effects of the history of the suspension.

When the aspect ratio of deformable objects is large, those are generally designated as fibers. Many
previous investigations of fiber dynamics, have focused on the dynamics of rigid fibers or rods \cite{Cox1971,Meunier1994}.
%while relatively little attention has been paid to the more complicated system of flexible fibers.
 Compared to the very large number of references related to particle suspensions, lower attention has been paid to the more complicated system of flexible fibers in a fluid.

Suspension of flexible fibers are encountered in the study of polymer dynamics \cite{Yamakawa1970,Yamanoi2010} whose rheology depends on the formation of networks and the occurrence of entanglement. The motion of fibers in
a viscous fluid has a strong effect on its bulk viscosity, microstructure, drainage rate, filtration ability, and flocculation properties. The dynamic response of such complex solutions is still an open issue while time-dependent structural changes of the dispersed fibers can dramatically modify the overall process (such as operation units in wood pulp and paper industry, flow molding techniques of composites, water purification).
Biological fibers such as DNA or actin filaments have also attracted many researches to understand the relation between
flexibility and physiological properties \cite{Jian1997}. 

Flexible fibers do not only passively respond to carrying flow gradients but can also be dynamically activated. 
Many of single cell micro-organisms that propel themselves in a fluid utilize a long flagellum tail connected to the cell body. 
Spermatozoa (and more generally one-armed swimmers) swim by propagating bending waves along their flagellum tail to generate a net translation using cyclic non-reciprocal strategy at low Reynolds number \cite{Purcell1977}. 
These natural swimmers have been modeled by artificial swimmers (joint microbeads)   actuated by an oscillating ambient
electric or magnetic field which opens breakthrough technologies for drug on-demand delivery in the human body \cite{Dreyfus2005}.

 %Many methods 
 %Lubrication  forces should in principle  prevents smooth spheres from touching, but 
 %in practice,  finite time stepping result in posible colision.
 %Because the computational cost for solving the ﬂow in the narrow gaps between closelyspaced particles is one the main challenging  difficult task different strategies have been
 %set-up to  
 %In reality, the surface of particles has some roughness and the bumps make physical contact

Many numerical methods have been proposed to tackle  elasto-hydrodynamic coupling between a fluid flow and deformable objects, i.e. the balance between viscous drag and elastic stresses.
Among those,  ``mesh-oriented'' approaches have the
ambition of solving a complete  continuum mechanics  description of the fluid/solid interaction, even though some approximations are mandatory to describe those at the fluid/solid interface. Without being all-comprehensive, one can cite immerse boundary methods (e.g. \cite{Fogelson:1988:FNM:56813.56816,Stockie_98,Zhu2002,Zhu2007}),  extended finite
elements  (e.g. \cite{Wagner2001}), penalty methods \cite{Glowinski1998181,Decoene2011}, particle-mesh Ewald methods \cite{Saintillan2005}, regularized Stokeslets \cite{Olson2013,Simons2014}, Force Coupling Method \cite{Yeo2010}.

In the specific context of low Reynolds number \emph{elastohydrodynamics} \cite{Wiggins1998}, difficulties arise  when numerically solving the dynamics of rigid objects since the time scale associated with elastic waves propagation within the solid can be similar to the viscous dissipation time-scale. In the context of self propelled objects the ratio of these time scales is called ``Sperm number''. When the Sperm number
is smaller or equal to one, the object temporal response is stiff, and requires small time steps to capture  fast deformation modes. In this regime, fluid/structure interaction effects are difficult to capture. 
%, even if  viscous diffusion is damped by creeping flow at low Reynolds number. 
One possible way to circumvent such difficulties is to use the knowledge of hydrodynamic interactions of simple objects in Stokes flow.

%the  interaction between objects in close contact. 
%For objects of regular shape and smooth surfaces such as sphere, ellipsoids, cylinders, etc.. close contact hydrodynamic interactions produces huge forces due to thin fluid film drainage, the so-called lubrication forces. This is also true when the objects are deformable. Computing    singular lubrication forces with a "mesh-oriented" approach involves many problems for very large velocity gradient necessitates extreme mesh refinement at high numerical cost. Evidently such cost is much more affordable for two dimensional meshes e.g (for boundary integral methods) but, it turns out that the resulting linear system might also become badly conditionned near contact \cite{Loewenberg_97}.

%In the context of non-deformable low Reynolds number flows, this

%Such difficulties can be circumvented by switching from numerical evaluation to analytically known lubrication forces close to contact, avoiding the cost of their numerical evaluation and providing
%very accurate descriptions \cite{Yeo2010,Swan2011}. Nevertheless, the use of lubrication force is only possible to tackle the close contact hydrodynamic interaction  between nearby objects of known shape.  Hence, there are compiling arguments to seek for a similar treatment of close interactions in the case of deformed structures described as flexible assembly of non-deformable objects. 

This strategy is the one pursued by the Bead Model (BM)  whose aim is to describe a complex deformable object by the flexible assembly of simple rigid ones.
 Such flexible assemblies are generally composed of beads (spheres or ellipsoids) interacting by some elastic and  repulsive forces, as well as with the surrounding fluid. 
For long elongated structures, alternative approaches to BM are indeed possible such as slender body approximation \cite{Tornberg2004, gaffney_11,Pozrikidis2011} or Resistive Force Theory \cite{Lauga2007,Coq2009,Gadelha2010a}.
% Nevertheless, such one-dimensional descriptions of elongated objects 
%can hardly take into account the finite size of the cross-section of the fibers and the hydrodynamic interactions with the flow or between the objects. 

One important advantage of BM which might explain their use among various communities (polymer Physics \cite{Gao2012,Jendrejack2002,Montesi2005,Yamanoi2011}, 
micro-swimmer modeling in bio-fluid mechanics \cite{Swan2011,Bilbao2013,Gauger2006,Majmudar2012}, flexible fiber in chemical engineering \cite{Lindstrom2007,Ross1997,Slowicka2012,Wang2012}), relies on their parametric versatility,  their ubiquitous character and their relative easy implementation. 
We provide a deeper, comparative and critical discussion about BM in Section \ref{BM}.  
 However, we would like to stress that the presented model is more clearly oriented toward micro-swimmer modeling than polymer dynamics.

One  should  also add that BM can be coupled to mesh-oriented approaches in order to  provide   accurate description of hydrodynamic interactions among large collection of deformable objects at moderate numerical cost  \cite{Majmudar2012}. 
Many authors only consider \emph{free drain}, i.e no Hydrodynamic Interactions (no HI),  \cite{Yamamoto1993, Skjetne1997, Ross1997, Schmid2000} or far field interactions associated with the Rotne-Prager-Yamakawa tensor \cite{Gauger2006, Manghi2006, Wada2006, Wajnryb2013}. 
This is supported by the fact that far-field hydrodynamic interactions  already provide  accurate predictions for the dynamics of a single flexible fiber when compared to experimental observations or numerical results.
 In order to illustrate the method we use, for convenience, the Rotne-Prager-Yamakawa tensor to model hydrodynamic interactions. 
We wish to stress here that this is not a limitation of the presented method, 
since the presented formulation holds for any mobility problem formulation. 
However, it turns out that for each configuration we tested, our model gave very good comparisons with other predictions, including those providing more accurate description of the hydrodynamic interactions.
 
The paper is organized as follows. First, we give a detailed presentation of the Bead Model for the simulation of flexible fibers. In this section, we propose a general formulation of kinematic constraints using the framework of Lagrange multipliers. 
This general formulation is used to present a new Bead Model, namely the Gears Model which surpasses existing models on numerical aspects. The second part of the paper is devoted to comparisons and validations of Bead Models for different configurations of flexible fibers (experiencing a flow or actuated filaments).

Finally, we conclude the paper by summarizing the achievements we obtain with our model and open new perspectives to this work.

\section{The Bead Model}
\label{BM}
\subsection{Detailed Review of previous Bead Models}
\label{sec:Detailed_review}

The Bead Model (BM) aims at discretizing any flexible object with interacting beads.
Interactions between beads break down into three categories: hydrodynamic interactions, elastic and kinematic constraint forces. 
Hydrodynamics of the whole object result from multibody hydrodynamic interactions between beads.
In the context of low Reynolds number, the relationship between stresses and velocities is linear.
Thus, the velocity of the assembly depends linearly on the forces and torques applied on each of its elements.
Elastic forces and torques are prescribed according to classical elasticity theory \cite{Landau1975} of flexible matter.
Constraint forces ensure that the beads obey any imposed kinematic constraint, e.g. fixed distance between adjacent particles.
All of these interactions can be treated separately as long as they are addressed in a consistent order.
The latter is the cornerstone which differentiates previous works in the literature from ours.
Numerous strategies have been employed to handle kinematic constraints.\\
 
\begin{figure}
\begin{centering}
\includegraphics[height=3cm]{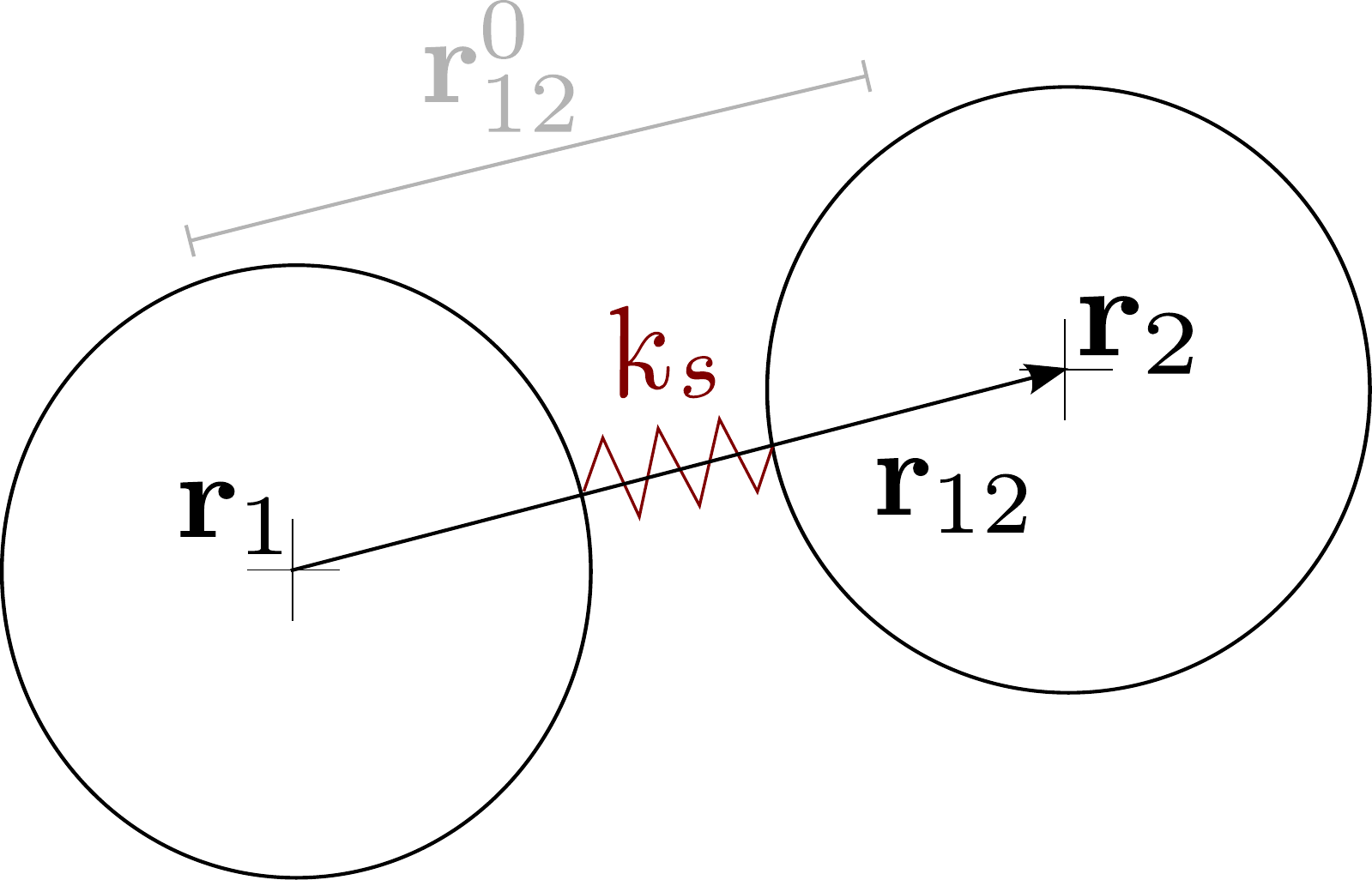}
\par\end{centering}
\caption{Spring model: linear spring to keep constant the inter-particle distance.}
\label{fig:Sketch_spring_model}
\end{figure}

\cite{Schlagberger2005, Gauger2006,  Wada2006, Yamanoi2011} and \cite{Slowicka2012} used a linear spring to model the resistance to stretching and compression without any constraint on the bead rotational motion (Fig. \ref{fig:Sketch_spring_model}).
The resulting stretching force reads:
\begin{equation}
 \mathbf{F}^{s}=-k_{s}(\mathbf{r}_{i,i+1}-\mathbf{r}^{0}_{i,i+1})
\end{equation}
where 
\begin{itemize}
 \item $k_{s}$ is the spring stiffness,
 \item $\mathbf{r}_{i,i+1} = \mathbf{r}_{i+1} - \mathbf{r}_{i}$ is the distance vector between two adjacent beads (for simplicity, equations and figures will be presented for beads $1$ and $2$ and can easily be generalized to beads $i$ and $i+1$),
 \item $\mathbf{r}_{1,2}^{0}$ is the vector corresponding to equilibrium.
\end{itemize}

However, regarding the connectivity constraint, the spring model is somehow approximate. A linear spring is prone to uncontrolled oscillations and the problem may become unstable. 
 Many other authors, among which \cite{Jendrejack2000,Jendrejack2002,Schroeder2004}, 
 thus use non-linear  spring models for a better description of polymer physics.   Nevertheless, the repulsive force stiffness has an important numerical cost in  time-stepping  as will be discussed in section \ref{time_step}. 
 Furthermore, unconstrained bead rotational motion leads to spurious hydrodynamic interactions and thus limits the range of applications for these BM.\\

\begin{figure}
\begin{centering}
\includegraphics[height=3cm]{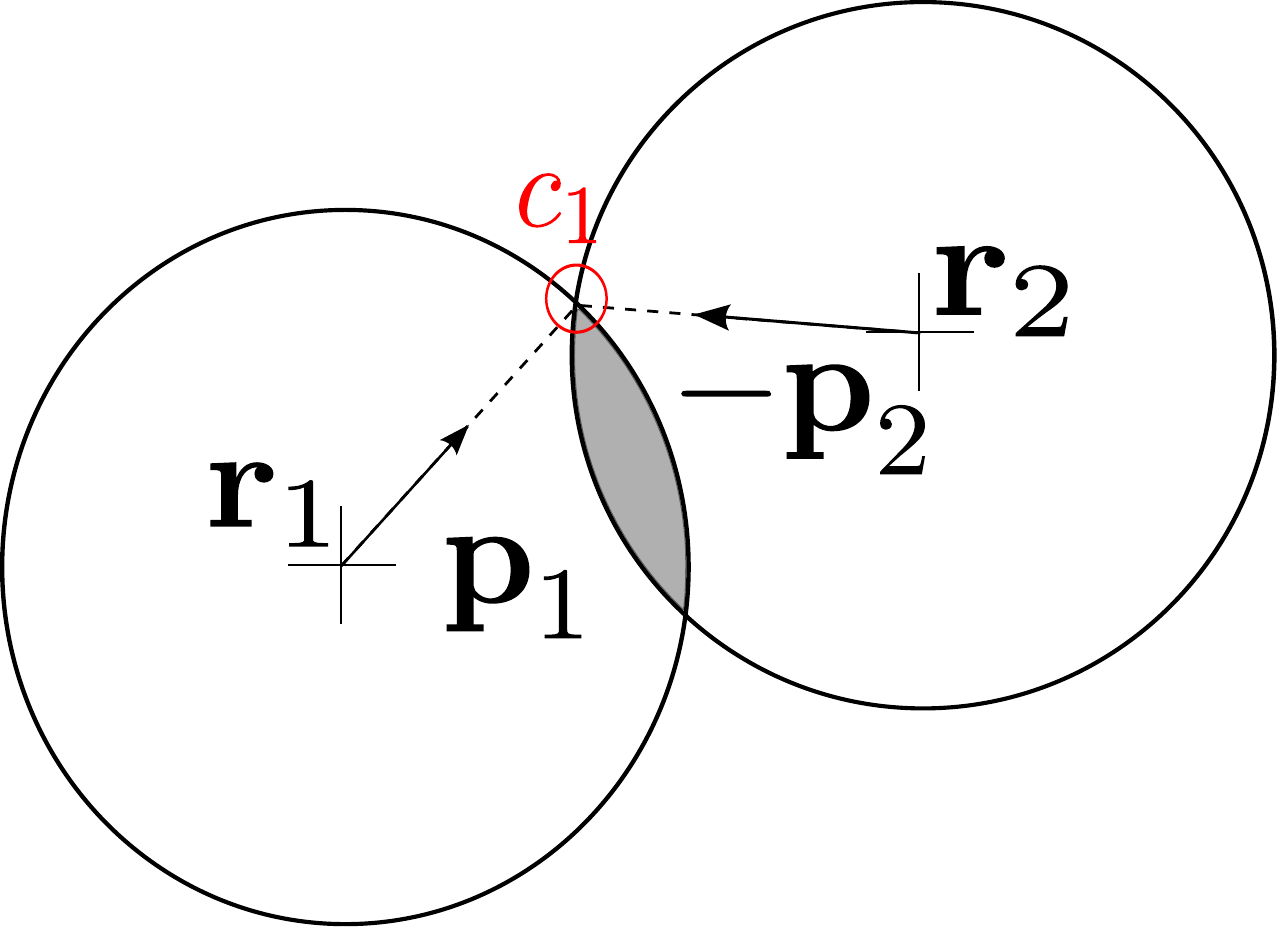}
\par\end{centering}
\caption{Joint Model: overlapping due to bending if no gap between beads.}
\label{fig:Sketch_joint_model_no_gap}
\end{figure}

Alternatively, \cite{Skjetne1997, Ross1997, Schmid2000, Qi2006} and \cite{Lindstrom2007} constrained the motion of the beads such that the contact point for each pair $c_i$ remains the same. 
While more representative of a flexible object, this approach exhibits two main drawbacks: 
\begin{enumerate}
 \item a gap between beads is necessary to allow the object to bend (see Fig. \ref{fig:Sketch_joint_model_no_gap}),
 \item it requires an additional center to center repulsive force, and thus more tuning numerical parameters to prevent overlapping between adjacent beads.
\end{enumerate}

\begin{figure}
\begin{centering}
\includegraphics[height=3cm]{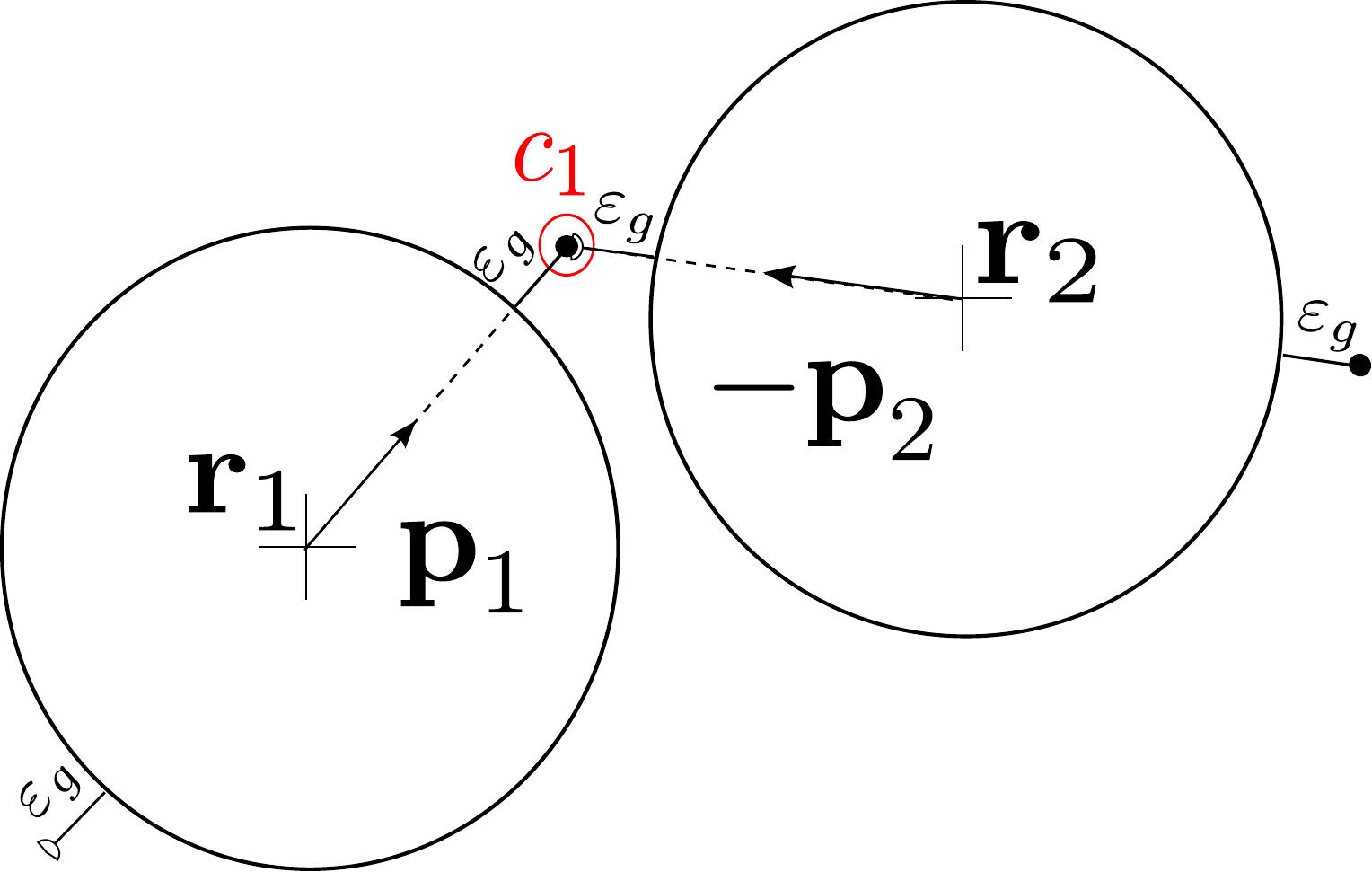}
\par\end{centering}
\caption{Joint Model: $c_1$ is separated by a gap $\varepsilon_g$ from the beads.}
\label{fig:Sketch_joint_model_gap}
\end{figure}

Consider two adjacent beads, with radius $a$, linked by a hinge $c_1$ (typically called ball and socket joint). 
The gap $\varepsilon_g$ defines the distance between the sphere surfaces and the joint (see Fig. \ref{fig:Sketch_joint_model_gap}).
Denote $\mathbf{p}_{i}$ the vector attached to bead $i$ pointing towards the next joint, i.e. the contact point $c_i$. 

\begin{figure}
\begin{centering}
\includegraphics[height=3cm]{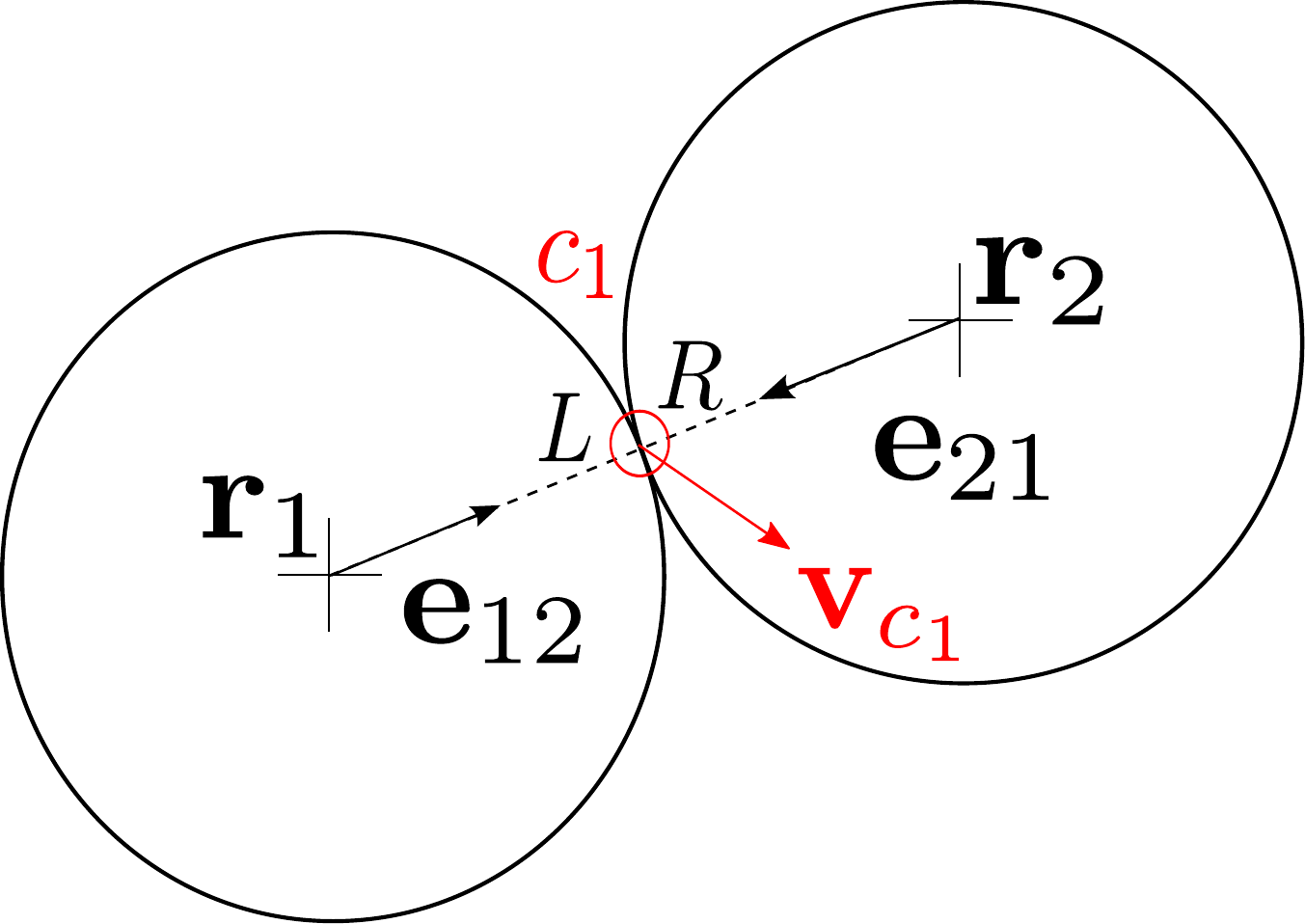}
\par\end{centering}
\caption{Gears Model: contact velocity must be the same for each bead (no-slip condition).}
\label{fig:Sketch_gears_model}
\end{figure}

The connectivity between two contiguous bodies writes:
\begin{equation}
 \left[\mathbf{r}_{1}+(a+\varepsilon_g)\mathbf{p}_{1}\right]-\left[\mathbf{r}_{2}-(a+\varepsilon_g)\mathbf{p}_{2}\right]=\mathbf{0}
\end{equation}
and its time derivative
\begin{equation}
 \left[\dot{\mathbf{r}}_{1}-(a+\varepsilon_g)\mathbf{p}_{1}\times\boldsymbol{\omega}_{1}\right]-\left[\dot{\mathbf{r}}_{2}+(a+\varepsilon_g)\mathbf{p}_{2}\times\boldsymbol{\omega}_{2}\right]=\mathbf{0}.
 \label{eq:Spherical_joint_kinematic}
\end{equation}
$\dot{\mathbf{r}}_{i}$ and $\boldsymbol{\omega}_{i}$ are the translational and rotational velocities of bead $i$.
The constraint forces and torques associated to \eqref{eq:Spherical_joint_kinematic} are obtained either by solving a linear system of equations involving beads velocities \cite{Schmid2000}, or by inserting \eqref{eq:Spherical_joint_kinematic} into the equations of motion when neglecting hydrodynamic interactions \cite{Skjetne1997, Ross1997}.\\
The gap width $2\varepsilon_g$ controls the maximum curvature $\kappa^{J}_{\max}$ allowed without overlapping. 
From the sine rule, one can derived the simple equation relating $\varepsilon_g$ and $\kappa^{J}_{\max}$
\begin{equation}
 \kappa^{J}_{\max} = \frac{\sqrt{1-\left(\frac{a}{a+\varepsilon_g}\right)^{2}}}{a}
\end{equation}
Once aware of these limitations, the gap  $\varepsilon_g$,  range and strength of the repulsive force should be prescribed depending on the problem to be addressed.\\
\cite{Keaveny2008} and \cite{Majmudar2012} proposed a more sophisticated Joint Model than those hitherto cited, using a full description of the links dynamics along the curvilinear abscissa. 
They derived a subtle constraint formulation which ensures that the tangent vector to the centerline is continuous and that the length of links remains constant.
These two works are worth mentioning since they avoid an empirical tuning of repulsive forces. Yet, \cite{Keaveny2008} computed the constraint forces and torques with an iterative penalty scheme instead of using an explicit formulation.\\

Finally, it is worth mentioning that the bead model proposed in \cite{Montesi2005} circumvents the
inextensibility difficulty by imposing constraints on the relative velocities of each successive segments, so that their relative distance is kept constant. 
Using bending potential, \cite{Montesi2005} permit overlap between beads with restoring torque (cf. Fig. \ref{fig:Sketch_joint_model_no_gap}). 
A Lagrangian multiplier formulation of tensile forces is also used in \cite{DoylePSShaqfehESG1997},
%where,  integrating the tensile gradient force  along the neutral fiber,  provides discrete tensile forces 
%oriented along each successive bead segment
which is equivalent to a prescribed equal distance between
successive beads. 
Again, inextensibility condition does not prevent bead overlapping due to bending in this formulation. 
 The computation of contact forces which is proposed in the following section \ref{section:Euler Lagrange}
generalizes the Lagrangian multiplier formulation of  \cite{Montesi2005} to generalized forces. 
Using more complex constraints involving both translational and angular velocities,
we show  that it is possible to accommodate both non-overlapping and inextensibility conditions without additional
repulsive forces (using the rolling no-slip contact with the gears model detailed in \ref{gearmodel}).
This proposed general formulation is also well suited for any type of kinematic boundary conditions as illustrated in Section \ref{subsec:Actuated_Filament}.

%An in-depth survey of the literature reveals that a generic model based on a well-established formalism could be useful to exploit efficiently the versatility and the simplicity of the Bead Model. 

\subsection{Generalized forces, virtual work principle and Lagrange multipliers }
\label{section:Euler Lagrange}

The model and formalism proposed in this article rely on earlier work in Analytical Mechanics and Robotics \cite{Nikravesh, Joshi2010}. 
The concept of generalized coordinates and constraints which has proven to be very useful in these contexts is described here. 
Generalized coordinates refer to  a set of parameters which uniquely describes the configuration of the system relative to some reference parameters (positions, angles,...). 
 For describing objects of complex shape, let us consider the position  ${ \bf{r}}_i$ of each bead $i\in \{1,N_b\}$  with associated orientation vector ${ \bf{p}}_i$ which is defined  by three Euler angles  $\bf{p}\equiv(\theta,\phi,\psi)$.
In the following,  any collection of vector population $({ \bf{r}}_1, ..{ \bf{r}}_i, ..{ \bf{r}}_{N_b} )\equiv { \bf{R}}$ will be capitalized, so that ${ \bf{R}}$ is  a vector in $\R^{3N_b}$. 
Hence the collection of orientation vectors ${ \bf{p}}_i$  will be denoted ${ \bf{P}}$, which is  a vector of  length $3{N_b}$, the collection of velocities $\frac{d{\bf{r}}_i}{dt} = { \bf{\dot{r}}}_i={\bf v}_i$, will be denoted ${ \bf{V}}$, the collection of angular velocity  ${\bf \dot p }_i\equiv{\bf  \omega }_i$ will be ${\bf  \Omega }$, the collection of forces  ${ \bf{f}}_i$, ${ \bf{F}}$, the collection of torques ${ \boldsymbol{\gamma}}_i$,  ${ \boldsymbol{\Gamma}}$. 
All ${ \bf{V}}$, ${\bf  \Omega }$, ${ \bf{F}}$ and ${ \bf{\Gamma}}$ are vectors in  $\R^{3N_b}$.

Let us then define some generalized coordinate ${ \bf q}_i$ for each bead, which is defined by ${ \bf q}_i\equiv ({ \bf{r}}_i,{ \bf{p}}_i)\equiv\{r_{1,i},r_{2,i},r_{3,i},\theta_i,\phi_i,\psi_i\}$ so that the collection of generalized positions  $({ \bf{q}}_1, ..{ \bf{q}}_i, ..{ \bf{q}}_{N_b} )\equiv { \bf{Q}}$ is  a vector in $\R^{6N_b}$. Generalized velocities are then defined by vectors ${ \bf \dot q}_i \equiv ({ \bf{v}}_i,{ \boldsymbol{\omega}}_i)$ with associated generalized collection of velocities ${  \bf{ \dot Q}}$.

Articulated systems are generically submitted to constraints which are either holonomic, non-holonomic or both \cite{Bailey2008}. 
Holonomic constraints do not depend on any kinematic parameter (i.e any translational or angular velocity) whereas  non-holonomic constraints do.
 
In the following we   consider  non-holonomic  linear kinematic constraints associated with generalized velocities of the form \cite{Greenwood1997}

%Holonomic constraints are relations between the generalized coordinates which can be expressed in the following form Let us now consider some non-holonomic constraints on the beads 
\begin{equation}
\label{non-holonomous_constraint}
{\bf{ \mathcal J}}  {  \bf{ \dot Q}} + {\bf B}  ={\bf 0},
\end{equation}
such that ${\bf \mathcal J}$ is a  $6N_b \times N_c$ matrix  and $\bf{B} $ 
is a vector of $N_c$ components. $N_c$ is the number of constraints acting on the $N_b$ beads. 
${\bf B}$ and ${\bf{ \mathcal J}}$ might depend (even non-linearly) on time $t$ and generalized positions ${\bf  Q}$, but do not depend  on any velocity of vector ${ \bf \dot Q}$, so that relation (\ref{non-holonomous_constraint}) is linear in ${ \bf \dot Q}$. 
In subsequent sections, we  provide  specific examples for which this class of constraints are useful. 
Here we describe, following  \cite{Greenwood1997, Nikravesh} how such constraints can be handled thanks to some generalized force that can be defined from Lagrange multipliers. 
The idea formulated to include constraints in the dynamics of articulated systems is to search additional forces which could permit to satisfy these constraints.  
First, one must rely on generalized forces ${\bf \mathfrak f}_i\equiv({\bf f}_i,{\boldsymbol{\gamma}}_i)$ which  include forces and torques acting on each bead, whose collection $({\bf \mathfrak f}_1, {\bf \mathfrak f}_i,..{\bf \mathfrak f}_{N_b})$ is denoted $\bf \mathfrak F$. 
Generalized forces are defined such that the  total work variation $ \delta W$ is the scalar product between them and the generalized coordinates variations ${\bf \delta Q}$ 
\begin{equation}
\label{genralyze_force}
{ \delta W}= {\bf \mathfrak F} \cdot {\bf \delta Q}={\bf F} \cdot { \bf \delta R} + {\bf \Gamma} \cdot {\bf \delta P},
\end{equation}
so that, on the right hand side of (\ref{genralyze_force}) one also gets  the translational and the rotational components of the work.
Then, the idea of virtual work principle is to search some virtual  displacement ${\bf \delta Q}$ that will generate no work, so that
\begin{equation}
\label{virtual_work}
{\bf \mathfrak F} \cdot {\bf \delta  Q}=0.
\end{equation}

At the same time, by rewriting \eqref{non-holonomous_constraint} in differential form
\begin{equation}
\label{non-holonomous_constraint diff}
{\bf{ \mathcal J}}  d{\bf Q} + {\bf B}dt  ={\bf 0},
\end{equation}  
admissible virtual displacements, i.e those satisfying constraints (\ref{non-holonomous_constraint diff}), should satisfy
\begin{equation}
\label{admissible_virtual_displacements}
{\bf \mathcal J }  {\bf \delta Q}={\bf 0}.
\end{equation}
Combining the $N_c$ constraints (\ref{admissible_virtual_displacements}) with 
(\ref{virtual_work}) is possible using any linear combination of these constraints. 
Such linear combination involves  $N_c$ parameters, the so-called Lagrange multipliers which are the components of  a vector  ${\bf \lambda}$ in $\R^{N_c}$.
Then from the difference between (\ref{virtual_work}) and the $N_c$  linear combination of (\ref{admissible_virtual_displacements}) one gets 
\begin{equation}
\label{diff}
\left( {\bf \mathfrak F}\, \, - \, \, {\bf \lambda} \cdot {\bf \mathcal J } \right) \cdot  {\bf \delta Q}= 0.
\end{equation}
Prescribing an adequate constraint force
\begin{equation}
\label{generalized_contact_force}
{\bf \mathfrak F}_c={\bf \lambda} \cdot {\bf \mathcal J },
\end{equation}

permits to satisfy the required equality for any virtual displacement.  
Hence, the constraints can be handled by forcing the dynamics with additional forces, the amplitude of which are given by Lagrange multipliers, yet to be found. 
Note also, that this first result implies that both translational forces and rotational torques associated with the $N_c$ constraints are both associated with the same Lagrange multipliers.

This formalism is particularly suitable for low Reynolds number flows for which translational and angular velocities are linearly related to forces and torques acting on beads  by  the mobility  matrix ${\bf M}$
\begin{equation}
\label{mobilite}
 \colvec{\bf{V}}{\bf{\Omega}} ={\bf M}  \colvec{{\bf F}}{\bf{\Gamma}} + 
\left(\begin{array}{c}\mathbf{V}^{\infty}\\
\boldsymbol{\Omega}^{\infty}\end{array}\right) + \mathbf{C}:\mathbf{E}^{\infty}.
\end{equation}
$\mathbf{V}^{\infty}=\left(\mathbf{v}^{\infty}_1,...,\mathbf{v}^{\infty}_{N_b}\right)$ and $\boldsymbol{\Omega}^{\infty}=\left(\boldsymbol{\omega}^{\infty}_1,...,\boldsymbol{\omega}^{\infty}_{N_b}\right)$ correspond to the ambient flow evaluated at the centers of mass $\mathbf{r}_i$.
$\mathbf{E}^{\infty}$ is the rate of strain $3 \times 3$ tensor of the ambient flow. 
$\mathbf{C}$ is a third rank tensor called the shear disturbance tensor, it relates the particles velocities and rotations to $\mathbf{E}^{\infty}$ \cite{Wajnryb2013}.
Matrix ${\bf M}$ (and tensor $\mathbf{C}$) can also be re-organized into a generalized mobility matrix ${\bf \mathcal M}$ (generalized tensor ${\bf \mathcal C}$ resp.) in order to define  the linear relation between the previously defined generalized velocity and generalized force
\begin{equation}
\label{mobilite_generalized}
{\bf \dot Q} ={\bf  \mathcal M}  {\bf \mathfrak F} + \mathcal{V}^{\infty} + \mathcal{C}:\mathbf{E}^{\infty},
\end{equation}
where $\mathcal{V}^{\infty}=\left(\mathbf{v}^{\infty}_1,\boldsymbol{\omega}^{\infty}_1,...,\mathbf{v}^{\infty}_{N_b},\boldsymbol{\omega}^{\infty}_{N_b} \right)$.
The explicit correspondence between  the classical matrix ${\bf M}$ and the hereby proposed generalized coordinate formulation ${\bf  \mathcal M}  $ is given in  \ref{appendix-sec1}.
Hence, as opposed to  the Euler-Lagrange formalism of classical mechanics, the dynamics of low Reynolds number flows does not involve any inertial contribution, and provide a simple linear relationship between forces and motion.
In this framework, it is then easy to handle constraints with generalized forces, because the total force will be the sum of the known hydrodynamic  forces ${\bf \mathfrak F}_h$, elastic forces ${\bf \mathfrak F}_e$,  inner forces associated to active fibers ${\bf \mathfrak F}_a$ and the hereby discussed and yet unknown contact forces  ${\bf \mathfrak F}_c$ to verify kinematic constraints
%\begin{equation}
\begin{eqnarray}
\label{total_force}
{\bf \mathfrak F} & = & {\bf \mathfrak F'}+{\bf \mathfrak F}_c, \,\,\,\,\,\rm{with}\\
{\bf \mathfrak F'} & = & {\bf \mathfrak F}_h+{\bf \mathfrak F}_e+{\bf \mathfrak F}_a.
\end{eqnarray}
%\end{equation}

Hence, if one is able to compute the Lagrange multipliers $\lambda$, the contact forces will provide the total force by linear superposition (\ref{total_force}), which gives the generalized velocities with (\ref{mobilite_generalized}). 
Now, let us show how  to compute the Lagrange multiplier vector. 
Since the generalized force is decomposed into known forces ${\bf \mathfrak F'}$  and unknown contact forces ${\bf \mathfrak F}_c={\bf \lambda} \cdot {\bf \mathcal J } $, relations (\ref{total_force}) and  (\ref{mobilite_generalized}) can be pooled together yielding
\begin{eqnarray}
\label{F_c}
{\bf  \mathcal M}{\bf \mathfrak F}_c={\bf  \mathcal M} {\bf \lambda} {\bf \mathcal J }& = & {\bf \dot Q}-{\bf  \mathcal M} {\bf \mathfrak F'} -\mathcal{V}^{\infty} - \mathcal{C}:\mathbf{E}^{\infty} .
\end{eqnarray}

So that, using (\ref{non-holonomous_constraint}), 

\begin{equation}
\label{Lagrange_multiplier}
{\bf \mathcal J } {\bf  \mathcal M} {\bf \mathcal J }^T{\bf \lambda} =- {\bf  B}-{\bf \mathcal J }\left( {\bf  \mathcal M} {\bf \mathfrak F'} + \mathcal{V}^{\infty} + \mathcal{C}:\mathbf{E}^{\infty} \right),
\end{equation}
one gets a simple linear system to solve for finding ${\bf \lambda}$, where ${\bf \mathcal J }^T$ stands for the transposition of  matrix ${\bf \mathcal J }$.

%- INTRODUCTION THEORIE GENERALE,\\
%- DEMONTRER LA RELATION ENTRE CONTRAINTES ET FORCES GENERALISEES,\\

%BIEN DIRE QUE:\\%
%
%-  NOS CONTRAINTES SONT NON-HOLONOMES CAR ON UTILISE LES VITESSES CAR PLUS SIMPLE CAR LINEAIRE CHEZ STOKES,\\
%-  PARCE QUE NOS CONTRAINTES SONT NON-HOLONOMES ON DOIT UTILISER TOUTES LES COORDONNEES GENERALISEES ET SUIVRE NIKRAVESH\\
%-  GARDER LA FORME GENERALE = JACOBIENNE*VITESSE = RHS\\

\subsection{The Gears Model}
\label{gearmodel}

The Euler-Lagrange formalism can be readily applied to any type of non-holonomic constraint such as \eqref{eq:Spherical_joint_kinematic}.
In the following, we propose an alternative model based on no-slip condition between the beads: the Gears Model. 
This constraint, first introduced in a Bead Model (BM) by \cite{Yamamoto1993}, conveniently avoid numerical tricks such as artificial gaps and repulsive forces.\\
However, \cite{Yamamoto1993} and \cite{Yamamoto1995} relied on to an iterative procedure to meet requirements.
Here, we use the Euler-Lagrange formalism to handle the kinematic constraints associated to the Gears Model.\\

Considering two adjacent beads (Fig. \ref{fig:Sketch_gears_model}), the velocity $\mathbf{v}_{c_1}$ at the contact point must be the same for each sphere:

\begin{equation}
 \mathbf{v}^1_{c_1} - \mathbf{v}^2_{c_1} = \mathbf{0}.
 \label{eq:No_Slip_1}
\end{equation}
$\mathbf{v}^L_{c_1}$ and $\mathbf{v}^R_{c_1}$  are respectively the rigid body velocity at the contact point on bead 1 and bead 2.
Denote $ \boldsymbol{\sigma}^1$ the vectorial no-slip constraint.  \eqref{eq:No_Slip_1} becomes

\begin{equation}
  \boldsymbol{\sigma}^{1}\left(\dot{\mathbf{r}}_{1},\boldsymbol{\omega}_{1},\dot{\mathbf{r}}_{2},\boldsymbol{\omega}_{2}\right)=\mathbf{0},
 \label{eq:No_Slip_2}
\end{equation}
i.e.
\begin{equation}
   \left[\dot{\mathbf{r}}_{1} - a\mathbf{e}_{12}\times\boldsymbol{\omega}_{1}\right] - \left[\dot{\mathbf{r}}_{2} - a\mathbf{e}_{21}\times\boldsymbol{\omega}_{2}\right]=\mathbf{0},
 \label{eq:No_Slip_3}
\end{equation}
where $\mathbf{e}_{12}$ is the unit vector connecting the center of bead 1, located at $\mathbf{r}_1$, to the center of bead 2, located at $\mathbf{r}_2$ ($\mathbf{e}_{12}=\mathbf{e}_{2}-\mathbf{e}_{1}$). The orientation $\mathbf{p}_i$ vector attached to bead $i$, is not necessary to describe the system.
Hence, from (\ref{eq:No_Slip_3}) one realises that 
$\boldsymbol{\sigma}^1$ is linear in translational and rotational velocities. Therefore equation \eqref{eq:No_Slip_2} can be reformulated as
\begin{equation}
  \boldsymbol{\sigma}^{1}\left({\bf \dot Q}\right)=\mathbf{J}^1{\bf \dot Q}=\mathbf{0}.
 \label{eq:No_Slip_4}
\end{equation}

where, ${\bf \dot Q}$ is the collection vector of generalized velocities of the two-bead assembly
\begin{equation}
  {\bf \dot Q}=\left[ \dot{\mathbf{r}}_{1},\, \boldsymbol{\omega}_{1},\, \dot{\mathbf{r}}_{2},\, \boldsymbol{\omega}_{2} \right]^T,
\end{equation}

$\mathbf{J}^{1}$ is the Jacobian matrix of $\boldsymbol{\sigma}^{1}$: 
\begin{equation}
  J_{kl}^{1}=\frac{\partial \sigma_{k}^{1}}{\partial Q_{l}},\,\, k=1,...,3,\,\, l=1,...,12,
\end{equation}

\begin{equation}
  \begin{array}{cll}
\mathbf{J}^{1} & = & \left[\begin{array}{cc}
\mathbf{J}_{1}^{1} & \mathbf{J}_{2}^{1}\end{array}\right]\\
\\ & = & \left[\begin{array}{cccc}
\mathbf{I}_{3} & -a\mathbf{e}_{12}^{\times} & -\mathbf{I}_{3} & a\mathbf{e}_{21}^{\times}\end{array}\right],\end{array}
\end{equation}

and
\begin{equation}
 \mathbf{e}^{\times}=\left(\begin{array}{ccc}
0 & -e_{3} & e_{2}\\
e_{3} & 0 & -e_{1}\\
-e_{2} & e_{1} & 0\end{array}\right).
\end{equation}

For an assembly of $N_{b}$ beads, $N_{b}-1$ no-slip vectorial constraints must be satisfied. 
The Gears Model (GM) total Jacobian matrix $\mathcal{J}^{GM}$ is block bi-diagonal and reads 
\begin{equation}
\label{J_def}
  \mathcal{J}^{GM}=\left(\begin{array}{ccccc}
  \mathbf{J}_{1}^{1} & \mathbf{J}_{2}^{1}\\
  & \mathbf{J}_{2}^{2} & \mathbf{J}_{3}^{2}\\
  &  & \ddots & \ddots\\
  &  &  & \mathbf{J}_{N_{b}-1}^{N_{b}-1} & \mathbf{J}_{N_{b}}^{N_{b}-1}\end{array}\right)
\end{equation}

where $\mathbf{J}_{\beta}^{\alpha}$ is the $3\times6$ Jacobian matrix of the vectorial constraint $\alpha$ for the bead $\beta$.

The kinematic constraints for the whole assembly then read
\begin{equation}
 \mathcal{J}^{GM} {\bf \dot Q}=\mathbf{0}.
\end{equation}

The associated generalized forces $\mathfrak{F}_c$ are obtained following Section \ref{section:Euler Lagrange}.\\

\subsection{Elastic forces and torques}

We are considering \emph{elastohydrodynamics} of homogeneous flexible and inextensible fibers. These objects experience bending torques and elastic forces to recover their equilibrium shape. Bending moments derivation and discretization are provided. Then, the role of bending moments and constraint forces is addressed in the force and torque balance for the assembly.

\subsubsection{Bending moments}
The bending moment of an elastic beam is provided by the constitutive law \cite{Landau1975, Bishop2004}
\begin{equation}
 \mathbf{m}(s)=K^{b}\mathbf{t}\times\frac{d\mathbf{t}}{ds},
 \label{eq:Bending_moment_tangent}
\end{equation}
where  $K^b(s)$ is the bending rigidity, {\bf $\mathbf{t}$ is the tangent vector along the beam centerline and } $s$ is the curvilinear abscissa.
Using the Frenet-Serret formula 
\begin{equation}
\frac{d\mathbf{t}}{ds}=\kappa\mathbf{n},
\end{equation}
the bending moment writes
\begin{equation}
 \mathbf{m}(s)=K^b\kappa\mathbf{b},
 \label{eq:Bending_moment_curv}
\end{equation}
where $\kappa(s)$ defines the local  curvature, $\mathbf{n}(s)$ and $\mathbf{b}(s)$ are the normal and binormal vectors of the Frenet-Serret frame.
When the link considered is not straight at rest, with an equilibrium curvature $\kappa^{eq}(s)$, \eqref{eq:Bending_moment_curv} is modified into
\begin{equation}
 \mathbf{m}(s)=K^b\left(\kappa-\kappa^{eq}\right)\mathbf{b}.
 \label{eq:Bending_moment_curv_intrinsic}
\end{equation}

\begin{figure}
\begin{centering}
\includegraphics[height=6.5cm]{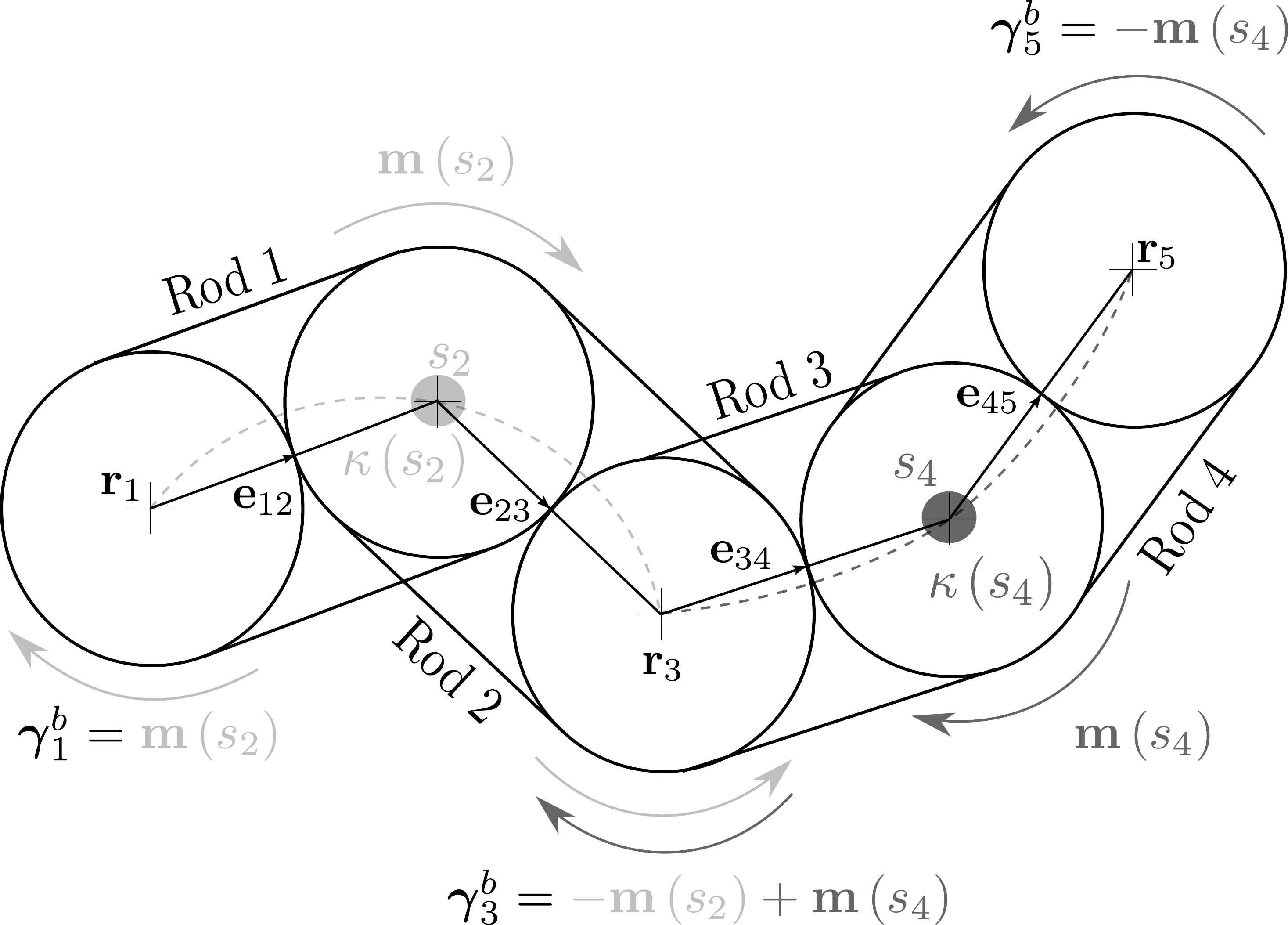}
\par\end{centering}
\caption{Beam discretization and bending torques computation of beads 1, 3 and 5.  
Remaining torques are accordingly obtained: $\boldsymbol{\gamma }_{2}^{b} = \mathbf{m}\left(s_{3}\right)$ and $\boldsymbol{\gamma}_{4}^{b} = -\mathbf{m}\left(s_{3}\right)$.}
\label{fig:Sketch_rods_bending}
\end{figure}

Here, the beam is discretized into $N_b-1$ rigid rods of length $l=2a$ (Cf Fig. \ref{fig:Sketch_rods_bending}).  
Inextensible rods are made up of two bond beads and linked together by a flexible joint with bending rigidity $K^b$.
Bending moments are evaluated at joint locations $s_i=(i-1)l$ for $i=2,...,N_b-1$, where $s_i$ correspond to the curvilinear abscissa of 
the mass center of the $i^{th}$ bead.

The bending torque on bead $i$ is then given by
\begin{equation}
 \boldsymbol{\gamma}_{i}^{b}=\mathbf{m}\left(s_{i+1}\right)-\mathbf{m}\left(s_{i-1}\right),
 \label{eq:Bending_torque_bead}
\end{equation}
with $\mathbf{m}\left(s_{i}\right) = K^b\kappa\left(s_{i}\right)\mathbf{b}\left(s_{i}\right)$.
See Fig. \ref{fig:Sketch_rods_bending} for the torque computation on a beam discretized with four rods.\\

The local curvature $\kappa\left(s_{i}\right)$ is approximated using the sine rule \cite{Lowe2003}
\begin{equation}
 %\kappa(s_{i})=\dfrac{2\sin\left(\cos^{-1}\left(\mathbf{e}_{i-1,i}\mathbf{.}\mathbf{e}_{i,i+1}\right)\right)}{\left\Vert \mathbf{r}_{i+1}-\mathbf{r}_{i-1}\right\Vert _{2}}  
\kappa(s_{i})=\dfrac{1}{a}\sqrt{\dfrac{1+\mathbf{e}_{i-1,i}\mathbf{.}\mathbf{e}_{i,i+1}}{2}}
 \label{eq:Sine_rule_curv}  
\end{equation}
where $\mathbf{e}_{i-1,i}$ is the unit vector connecting the center of mass of bead $i-1$ to the center of mass of bead $i$.
This elementary geometric law provides the radius of curvature $R(s_i) = 1/\kappa(s_i)$ of the circle circumscribing neighbouring bead centers $\mathbf{r}_{i-1}$, $\mathbf{r}_{i}$ and $\mathbf{r}_{i+1}$.

 A more general version of the discrete curvature proposed in \cite{Fauci1988} can also be used in the case of three dimensional
motion. In that case, the curvature of the fiber is discretized as in \cite{Fauci1988}
\begin{equation}
 \boldsymbol{\kappa}\left(s_{i}\right)=\dfrac{\mathbf{e}_{i-1,i}\times\mathbf{e}_{i,i+1}}{2a},
 \label{eq:curv_3D}
\end{equation}
where, again, $\mathbf{e}_{i-1,i}$ is the unit vector connecting the center of mass of bead $i-1$ to the center of mass of bead $i$. The bending moment reads
\begin{equation}
 \mathbf{m}\left(s_{i}\right)=K^b  \boldsymbol{\kappa} \left(s_{i}\right).
\end{equation}\\
To include the effect of torsional twisting about the axis of the fiber, one would have to compute the relative orientation between the frames of reference attached to the beads using Euler angles \cite{Keaveny2008} (see Section \ref{section:Euler Lagrange}) or unit quaternions as in \cite{Schmid2000}.
This would provide the rate of change of the twist angle along the fiber centerline and thus the twisting torque acting on each bead.
In the following, only bending effects are considered.

\subsubsection{Force and moment acting on each bead}

The Gears Model proposed in this paper does not need to consider gaps to allow bending. 
$\mathfrak{F}_c$ also ensures the connectivity condition and circumvent the use of repulsive forces as distances between adjacent bead surfaces remain constant. 
More specifically, the tangential components of the force  $\mathbf{F}_c $, which is only one part of the generalized force $\mathfrak{F}_c$, acts as tensile force.

 For each bead $i$, contact forces applied from bead  $i$ to bead
 $i+1$  at contact point $c_{i}$ between bead  $i$ and
 $i+1$  (Figure \ref{fig:Sketch_gears_model} for two beads) is denoted $\mathbf{f}_{c_i}$. From Newton third law at contact point  $c_{i}$,
 the contact force applied to bead $i$ from bead $i+1$ is obviously 
 $- \mathbf{f}_{c_i}$. Total force acting on bead $i$ from contact, and hydrodynamic forces $\mathbf{f}^h_{i}$ reads

 \begin{equation}
  \label{total_force_bead_i}
  \mathbf{f}_{i}=\mathbf{f}_{c_{i-1}}-\mathbf{f}_{c_{i}}+\mathbf{f}^h_{i}
\end{equation}
Similarly,  the contact force $\mathbf{f}_{c_i}$ at point $c_{i}$  produces a  moment ${ \bf m}_{c_i}=a {\bf t }_i \times { \bf f}_{c_i}$  associated with  local tangent  vector  ${\bf t}_i=\mathbf{e}_{i,i+1}$ and distance $a$ to the neutral fiber at point $c_{i}$. 
Total moment acting on bead $i$ from contact points moments, elastic and hydrodynamic torques are then 
\begin{equation}
\label{total_moment}
{ \boldsymbol{\gamma}}_{i}=\mathbf{m}_{c_{i-1}}-\mathbf{m}_{c_{i}}+ \boldsymbol{\gamma}^b_{i}+\boldsymbol{\gamma}^h_{i}.
\end{equation}
The contribution of contact force and contact moment acting on bead $i$   exactly equals
the contribution of the generalized contact force. Indeed, using the kinematic constraints Jacobian (\ref{J_def}) in (\ref{generalized_contact_force}), and computing the force and torque contributions, one exactly
recovers the first and the second contributions of the right-hand-side of (\ref{total_force_bead_i}) and (\ref{total_moment}). In  \ref{appendix-sec2}, we also show that this model is consistent with classical formulation for slender body force and moment balance when the bead radius tends to zero.

\subsection{Hydrodynamic coupling}
\label{section:Hydro_coupling}

Moving objects (rigid or flexible fibers) in  a viscous fluid experience hydrodynamic forcing. 
The interactions are mediated by  the fluid flow perturbations which can alter the motion and  the deformation of the fibers in a moderately concentrated suspension. 
The existence of hydrodynamic interactions has also  an effect on  a single fiber dynamics while different parts of the fiber can respond to the ambient flow but also to local flow perturbations related to the fiber deformation. 
Resistive Force Theory (RFT) can be used to estimate the fiber response to a given flow assuming that the fiber is modeled by a large series of slender objects \cite{Coq2010,Lauga2007}. 
Slender body theory has also been used \cite{Tornberg2004,Li2013} to relate local balance of drag forces with the filament forces upon the fluid resulting in a dynamical system to model the deformation of the fiber centerline. 
This model provided interesting results on the stretch-coil transition of fibers in vortical flows.

In our beads model, the fiber is composed of spherical particles to account for the finite width of its cross-section. 
The hydrodynamic interactions are provided through the solution of the mobility problem which relates forces, torques to the translational and rotational velocities of the beads. 
This many-body problem is non-linear  in the instantaneous positions of all particles of the system. 
 Approximate solutions of this complex mathematical problem can be achieved by limiting the mobility matrices to their leading order. 
The simplest model is called free drain as the mobility matrix is assumed to be diagonal neglecting the HI with neighbouring spheres. 
 Pairwise interactions are required to account for anisotropic drag effects within the beads composing the fiber. 
The Rotne-Prager-Yamakawa (RPY) approximation is one of the most commonly used methods of including hydrodynamic interactions. 
This widely used approach has been recently updated by Wajnryb et al. \cite{Wajnryb2013} for the RPY translational and rotational degrees of
freedom, as well as for the shear disturbance tensor $\mathbf{C}$ which gives the response of the particles to an external shear flow  \eqref{mobilite}.

\subsection{Numerical implementation}

\subsubsection{Integration scheme and algorithm}
\label{sec:Algo}

The kinematics of the constrained system results from the superposition of individual bead motions.
Positions are obtained from the temporal integration of the  equation of motion with a third order Adams-Bashforth scheme
\begin{equation}
 \dfrac{d\mathbf{r}_{i}}{dt}=\mathbf{v}_i,
 \label{eq:Lag_pos_velocities}
\end{equation}

where $\mathbf{r}_i$, $\mathbf{v}_i$ are the position and translational velocity of bead $i$.

The time step  $\Delta t$ used to integrate \eqref{eq:Lag_pos_velocities} is fixed by the characteristic bending time \cite{Lindstrom2007}
\begin{equation}
 \Delta t<\dfrac{\mu (2a)^{4}}{ K^{b}}.
\label{eq:calc_dt}
\end{equation}
where $\mu$ is the suspending fluid viscosity.

The evaluation of bead interactions must follow a specific order. 
Elastic and active forces can be computed in any order.
Constraint forces and torques must be estimated afterwards as they depend on $\mathfrak{F}'$.
Then velocities and rotations are obtained from the mobility relation. And finally, bead positions are updated.\\

\begin{itemize}
 \item Initialization: positions $\mathbf{r}_i(0)$,
 \item Time Loop
 \begin{enumerate}
  \item Evaluate mobility matrix $\mathcal{M}\left(\mathbf{Q}\right)$ and $\mathcal{C}:\mathbf{E}^{\infty}$ (see Section \ref{section:Hydro_coupling}),
  \item Calculate local curvatures \eqref{eq:Sine_rule_curv} and bending torques $\boldsymbol{\gamma}_i^b$  \eqref{eq:Bending_torque_bead} to get $\mathfrak{F}_e$,
  \item Add active forcing $\mathfrak{F}_a$ and ambient velocity $\mathcal{V}^{\infty}$ if any,
  \item Compute the Jacobian matrix associated with non-holonomic constraints $\mathcal{J}\left(\mathbf{Q}\right)$,
  \item Solve \eqref{Lagrange_multiplier} to get the constraint forces $\mathfrak{F}_c=\lambda\mathcal{J}$,
  \item Sum all the forcing terms $\mathfrak{F} = \mathfrak{F}_e + \mathfrak{F}_a + \mathfrak{F}_c$,
  \item Apply mobility relation \eqref{mobilite_generalized} to obtain the bead velocities $\dot{\mathbf{Q}}$,
  \item Integrate  \eqref{eq:Lag_pos_velocities} to get the new bead positions.
 \end{enumerate}
\end{itemize}

\subsubsection{Implementation of the Joint Model}
\label{Implementation_joint}

To provide a comprehensive comparison with previous works, we exploit the flexibility of the Euler-Lagrange formalism to implement the Joint Model as described in \cite{Skjetne1997} supplemented with hydrodynamic interactions. The joint constraint for two neighbouring beads reads 

\begin{equation}
 \left[\dot{\mathbf{r}}_{i}-(a+\varepsilon_g)\mathbf{p}_{i}\times\boldsymbol{\omega}_{i}\right]-\left[\dot{\mathbf{r}}_{i+1}+(a+\varepsilon_g)\mathbf{p}_{i+1}\times\boldsymbol{\omega}_{i+1}\right]=\mathbf{0}.
 \label{eq:Spherical_joint_kinematic2}
\end{equation}

Using the Euler-Lagrange formalism, \eqref{eq:Spherical_joint_kinematic2} is reformulated with the Joint Model (JM) Jacobian matrix

\begin{equation}
 \mathcal{J}^{JM} {\bf \dot Q}=\mathbf{0},
\end{equation}

where $ \mathcal{J}^{JM}$ has the same structure as in  \eqref{J_def} and 

\begin{equation}
  \begin{array}{cll}
\mathbf{J}^{i} & = & \left[\begin{array}{cc}
\mathbf{J}_{1}^{i} & \mathbf{J}_{2}^{i}\end{array}\right]\\
\\ & = & \left[\begin{array}{cccc}
\mathbf{I}_{3} & -(a+\varepsilon_g)\mathbf{p}_{i}^{\times} & -\mathbf{I}_{3} & -(a+\varepsilon_g)\mathbf{p}_{i+1}^{\times}\end{array}\right].\end{array}
\end{equation}

Accordingly, the corresponding set of forces and torques $\mathfrak{F}_c$ are obtained from Section \ref{section:Euler Lagrange}.
As mentioned in Section \ref{sec:Detailed_review}, such formulation does not prevent beads from overlapping when bending occurs.
A repulsive force $\mathbf{F}^r$ is added according to \cite{Lindstrom2007} (the force profile proposed by \cite{Skjetne1997} is very stiff, thus very constraining for the time step):

\begin{equation}
 \mathbf{F}_{ij}^{r}=\begin{cases}
-F_{0}\exp\left(-\dfrac{d_{ij}+\delta_{D}}{d_{0}}\right)\mathbf{e}_{ij}, & d_{ij}\leq-\delta_{D},\\
-F_{0}\left(\dfrac{1}{2}-\dfrac{d_{ij}}{2\delta_{D}}\right)\mathbf{e}_{ij}, & -\delta_{D}<d_{ij}\leq\delta_{D},\\
\mathbf{0}, & r_{ij}>\delta_{D}.\end{cases}
\end{equation}

$\delta_D$ is an artificial surface roughness, $d_{ij}$ is the surface to surface distance. 
$d_{ij}<0$ indicates overlapping between beads $i$ and $j$.
$d_0$ is a numerical damping distance which has to be tuned to prevent overlapping.
$F_0$ is the repulsive force scale chosen in order to avoid numerical instabilities. 
To deal with this issue, \cite{Lindstrom2007} proposed to evaluate $F_0$ from bending and viscous stresses.
A slight modification of their formula for inertialess particles yields

\begin{equation}
F_{0}=C_{1}6\pi\mu L\left(\overline{\mathbf{v^{\infty}}}-\overline{\mathbf{v}}\right)+C_{2}\sqrt{\dfrac{\overline{K^{b}}E^{b}}{L^{3}}},
\end{equation}

the bar denotes the average over the constitutive beads or joints where $C_{1}$ and $C_{2}$ are adjustable constants.
$E^b$ is the bending energy 

\begin{equation}
E^{b}=\sum_{i=1}^{N_{b}-1}K^{b}\left(\kappa(s_{i})-\kappa^{eq}(s_{i})\right)^{2},
\end{equation}

Bending moments are evaluated at the joint locations $s^J_i =(a+\varepsilon_g) + (i-1)\times2(a+\varepsilon_g), \ i=1,...,N_b-1$.
Joint curvature is given by 

\begin{equation}
\kappa(s^J_{i})=\dfrac{2}{a+\varepsilon_g}\sqrt{\dfrac{1+\mathbf{p}_{i} . \mathbf{p}_{i+1}}{2}},
 \label{eq:Sine_rule_curv_joint}  
\end{equation}

Similarly to  \eqref{eq:Bending_torque_bead}, bending torque on bead $i$ is 

\begin{equation}
 \boldsymbol{\gamma}_{i}^{b}=\mathbf{m}\left(s^J_{i}\right)-\mathbf{m}\left(s^J_{i-1}\right).
 \label{eq:Bending_torque_bead_joint}
\end{equation}

Bead orientation $\mathbf{p}_i$ is integrated with a third order Adams-Bashforth scheme
\begin{equation}
 \dfrac{d\mathbf{p}_{i}}{dt}=\boldsymbol{\omega}_{i}\times\mathbf{p}_{i}.
 \label{eq:eq_lag_orient}
\end{equation}

The procedure is similar to the Gears Model. $\mathbf{p}_{i}$ are initialized together with the positions.
The repulsive force $\mathbf{F}^r$ is added to $\mathfrak{F}'$ and  can be computed between step 1 and 5 of the aforementioned algorithm.
Time integration of \eqref{eq:eq_lag_orient} is performed at step 8.

\subsubsection{Constraints and numerical stability}
\label{time_step}
At each time step, the error on kinematic constraints $\epsilon$ is evaluated, after application of the mobility relation  \eqref{mobilite_generalized}, between step 7 and step 8: 
\begin{equation}
 \epsilon^{GM} (t) = \left\Vert \mathcal{J}^{GM}\dot{\mathbf{Q}}  \right\Vert_2  = \left(\sum_{i=1}^{N_{b}-1}\left(\mathbf{v}_{c_{i}}^{L}-\mathbf{v}_{c_{i}}^{R}\right)^{2}\right)^{1/2}
\end{equation}

for the Gears Model, and

\begin{equation}
 \epsilon^{JM} (t) = \left\Vert \mathcal{J}^{JM}\dot{\mathbf{Q}}  \right\Vert_2 
\end{equation}

for the Joint Model.\\

To verify the robustness of both models and Lagrange formulation, a numerical study is carried out on a stiff configuration.\\
A fiber of aspect ratio $r_p = 10$ with bending ratio $BR=0.01$ is initially aligned with a shear flow of magnitude $\dot{\gamma}=5s^{-1}$.
 For this aspect ratio,  $N_b=10$ beads are used to model the fiber with the  Gears Model.\\

Joint Model involves additional items to be fixed.  $N_b = 9$ spheres are separated by a gap width $2\varepsilon_g = 0.25a$.
 The repulsive force is activated when the surface to surface distance $d_{ij}$ reaches the artificial surface roughness $\delta_D = 2(a+\varepsilon_g)/10$.
 The remaining coefficients are set to reduce numerical instabilities without affecting the Physics of the system: $d_0 = (a+\varepsilon_g)/4 $, $C_1 = 5$ and $C_2=0.5$.\\

\begin{figure}
\begin{centering}
\includegraphics[width=10cm]{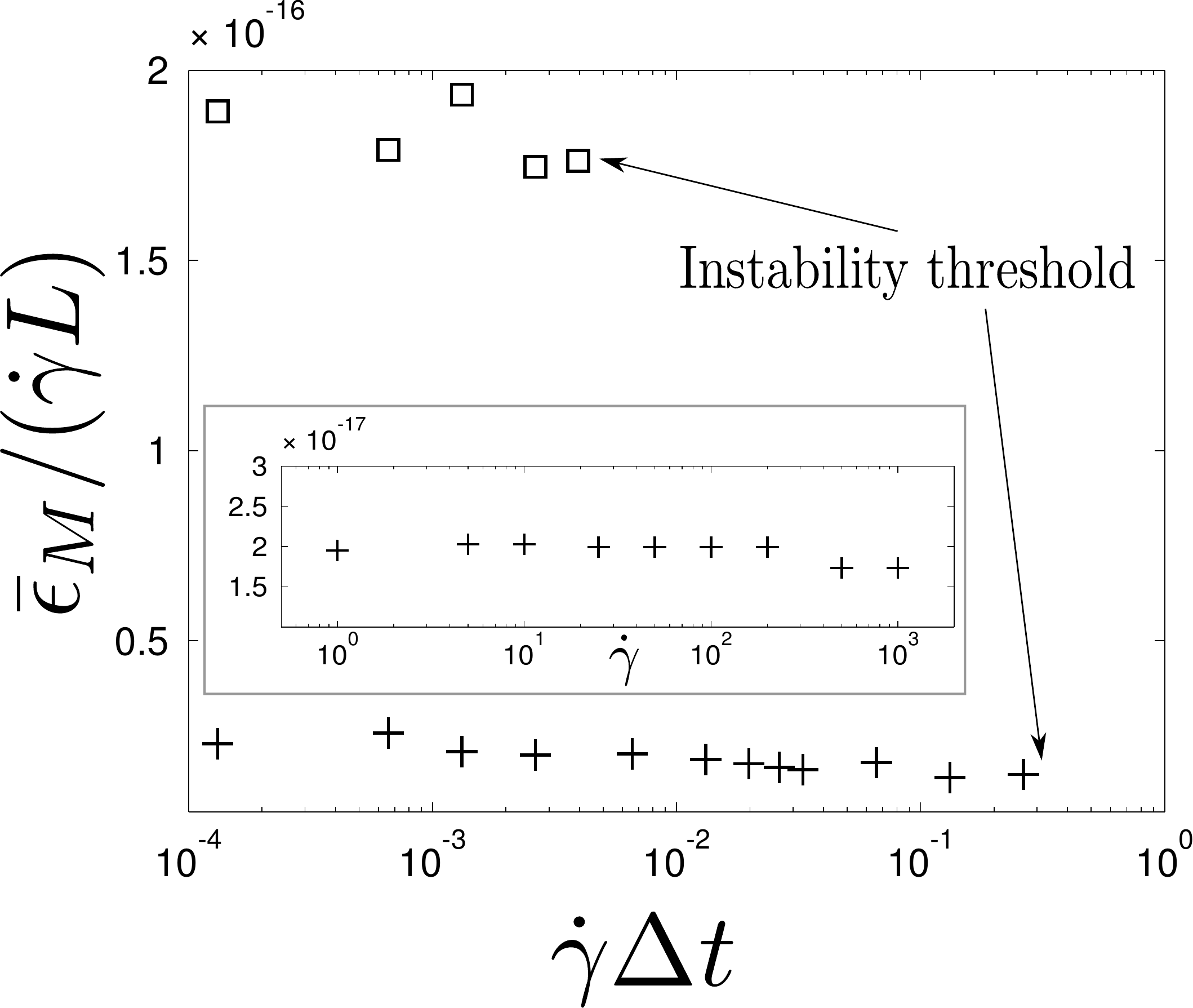}
\par\end{centering}
\caption{Dependence of the constraints $\bar{\epsilon}_M/\dot{\gamma}L$ on the time step $\dot{\gamma}\Delta t$, \symbol{}{\plus}{-5}{0}{black}{black}: Gears Model, \symbol{}{\ssquareb}{-5}{0}{black}{black}: Joint Model. Inset: $\bar{\epsilon}_M/\dot{\gamma}L$ with the Gears Model for a fixed time step given by \eqref{eq:calc_dt} for different values of $\dot{\gamma}$.}
\label{fig:NoSlip_dt}
\end{figure}

Fig. \ref{fig:NoSlip_dt} shows the evolution of the maximal mean deviation from the no-slip/joint constraint $\bar{\epsilon}_M = \underset{t}{\max} \ \epsilon (t)/(N_b-1) $  normalized with the maximal shear velocity $\dot{\gamma}L$ depending on the dimensionless time step $\dot{\gamma}\Delta t$.
First, one can observe that for both Joint and Gears models, $\bar{\epsilon}_M/\dot{\gamma}L$ weakly depends on $\dot{\gamma}\Delta t$ and the resulting motion of the beads complies very precisely with the set of constraints, within a tolerance close to unit roundoff ($<2.10^{-16}$).
Secondly, Joint Model is unstable for time steps 100 times smaller than Gears Model.
The onset for numerical instability indicates that the repulsive force stiffness dominates over bending, thus dictating and restricting the time step.\\
As a comparison, \cite{Lindstrom2007} matched connectivity constraints within $1\%$ error for each fiber segment. 
To do so, they had to use an iterative scheme reducing the time step by $1/3$ each iteration to meet requirements and limit overlapping between adjacent segments.
For similar results, a stiff configuration, such as the sheared fiber, is therefore more efficiently simulated with the Gears Model.\\
Thirdly, inset of Fig. \ref{fig:NoSlip_dt} shows that, for a given time step, the Gears Model constraints $\bar{\epsilon}_M/\dot{\gamma}L$ are satisfied whatever the shear magnitude. 
Hence, \eqref{eq:calc_dt} ensures unconditionally numerical stability as bending is the only limiting effect for the Gears Model.

Hence, the robustness of the Euler-Lagrange formalism and the numerical integration we chose 
provide a strong support to the Gears Model over the Joint Model.\\
 
 As a final remark to this section, it is important to  mention that the numerical cost of the proposed  
method strongly depends on the  choice for the mobility matrix computation, as usual for bead models. If the mobility matrix 
is computed taking into account full hydrodynamic interactions with Stokesian Dynamics, most of the numerical cost will come 
from its evaluation in this case. This limitation could be overcame  using more sophisticated methods
such as Accelerated  Stokesian Dynamics \cite{Sierou2001a} or Force Coupling Method \cite{Yeo2010}.
  Moreover, when considering Rotne-Prager-Yamakawa mobility matrix, 
 its cost only requires the  evaluation  of $O\left((6N_b)^2 \right)$ terms.  
 Furthermore, the main algorithmic  complexity of  bead models does not  come from the time integration of the bead positions which only 
requires a matrix-vector multiplication \eqref{mobilite_generalized} 
%of the mobility matrix onto the forces 
at a  $O\left((6N_b)^2\right)$ cost. 
Fast-multipole formulation of a  Rotne-Prager-Yamakawa matrix can even provide  a $O\left(6N_b\right)$ cost for such matrix-vector multiplication  \cite{Liang:2013:FMM:2405837.2405895}.

The main numerical cost indeed comes from the inversion of the contact forces problem 
\eqref{Lagrange_multiplier}. It is worth noting that this linear  problem is $N_c \times N_c$ which is slightly different from
 $N_b \times N_b$, but of the same order. Furthermore, problem \eqref{Lagrange_multiplier} gives a direct,
single step procedure to compute the contact forces, as opposed to previous other attemps \cite{Yamamoto1993,Lindstrom2007,Keaveny2008} which required
iterative procedures to meet forces requirements,
involving the mobility matrix inversion at each iteration.   
The cost for the inversion of \eqref{Lagrange_multiplier}  lies  in-between $O(N_c^2)$ and $O(N_c^3)$ depending on the inversion method. 
%It might turn out that problem \eqref{Lagrange_multiplier} associated  with band matrix ${\bf \mathcal J }$ and a Rotne-Prager-Yamakawa mobility matrix could be obtained in an efficient way.

\section{Validations}

\subsection{Jeffery Orbits of rigid fibers}
Much of our current understanding of the behavior of fibers experiencing a shear flow has come from the work of Jeffery \cite{Jeffery1922} who derived the equation for the motion of an ellipsoidal particle in Stokes flow. The same equation can be used for the motion of an axisymmetric particle by using an equivalent ellipsoidal aspect ratio. Rigid fibers can be approximated by elongated prolate ellipsoids. An isolated fiber in simple shear flow rotates in a periodic orbit while the center of mass simply translates in the flow (no migration across streamlines). The period $T$  \eqref{eq:Jeff_T} is a function of the aspect ratio of the fiber and the flow shear rate while the orbit depends on the initial orientation of the object relative to the shear plane

\begin{equation}
 \mbox{T}=\dfrac{2\pi (r_e + 1/r_e)}{\dot{\gamma}}.
 \label{eq:Jeff_T}
\end{equation}

$\dot{\gamma}$ is the shear rate of the carrying flow. $r_e$ is the equivalent ellipsoidal aspect ratio which is related to the fiber aspect ratio $r_p$ (length of the fiber over diameter of the cross-section which turns out to $r_p = N_b$ with $N_b$ beads). The fiber is initially placed in the plane of shear and is composed on $N_b$ beads. No gaps between beads is required in the Joint Model because the fiber is rigid and flexibility deformations are negligible. 
We have compared the results with two relations for $r_e$: Cox \cite{Cox1971} 

\begin{equation}
 r_e=\dfrac{1.24 r_p}{\sqrt{\ln(r_p)}},
 \label{eq:Cox_re}
\end{equation}

and Larson \cite{Larson1999}

\begin{equation}
 r_e=0.7 r_p
 \label{eq:Larson_re}.
\end{equation}

\begin{figure}
\begin{centering}
\includegraphics[height=7.5cm]{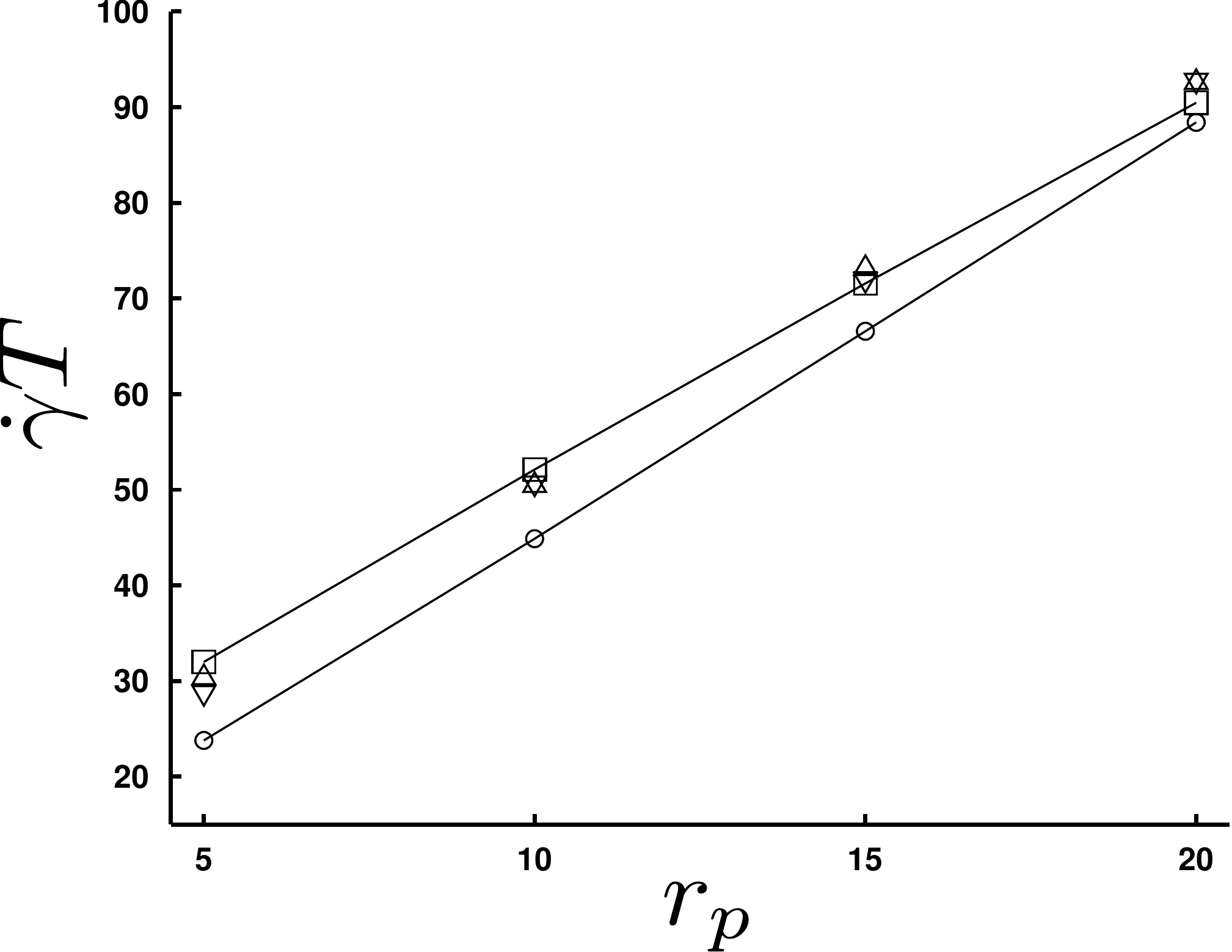}
\par\end{centering}
\caption{
Tumbling period $T$ depending on fiber aspect ratio $r_p$.
\symbol{\solid}{\circle}{22}{0}{black}{black} : theoretical law  \eqref{eq:Jeff_T} with $r_e$ given by \eqref{eq:Larson_re}, 
\symbol{\solid}{\ssquareb}{22}{1}{black}{black} : theoretical law  \eqref{eq:Jeff_T} with $r_e$ given by \eqref{eq:Cox_re},  
\symbol{}{\trianup}{-5}{2}{black}{black}: Gears Model, 
\symbol{}{\bigtriandown}{-5}{0}{black}{black}: Joint Model.
}
\label{fig:Jeffery}
\end{figure}

This classic and simple test case has been successfully validated in \cite{Yamamoto1993,Skjetne1997,Yamanoi2011}. Both the Joint and Gears models give a correct prediction of the period of Jeffery orbits (Fig. \ref{fig:Jeffery}). The scaled period $\dot{\gamma}T$ of simulations remains within the two evolutions based on equations \eqref{eq:Cox_re} and \eqref{eq:Larson_re}. We have tried to compare with the linear spring model proposed by Gauger and Stark \cite{Gauger2006} (and used by Slowicka et al. \cite{Slowicka2012} with a more detailed formulation of hydrodynamic interactions). In this latter model, there is no constraint on the rotation of beads and the simulations failed to reproduce Jeffery orbits (the fiber does not flip over the axis parallel to the flow). 

\subsection{Flexible fiber in a shear flow}

The motion of flexible fibers in a shear flow is essential in paper making or composite processing. 
Prediction and control of fiber orientations and positions are of particular interest in the study of flocks disintegration.
Many models have been designed to predict fiber dynamics and much experimental work has been conducted. 
The wide variety of fiber behaviors depends on the ratio of bending stresses over shear stress, which is quantified by a dimensionless number, the bending ratio BR \cite{Forgacs1959a, Schmid2000}
\begin{equation}
 \mbox{BR}=\dfrac{E(\ln2r_{e}-1.5)}{\mu\dot{\gamma}2r_{p}^{4}}
\end{equation}

$E$ is the material Young's Modulus and $\mu$ is the suspending fluid viscosity.\\
In the following, we investigate the response of the Gears Model with known results on flexible fiber dynamics.

\subsubsection{Effect of permanent deformation}
\begin{figure}
\begin{centering}
\subfloat[]{\label{fig:Snap_buckling_no_int_curv} \includegraphics[height=6.5cm]{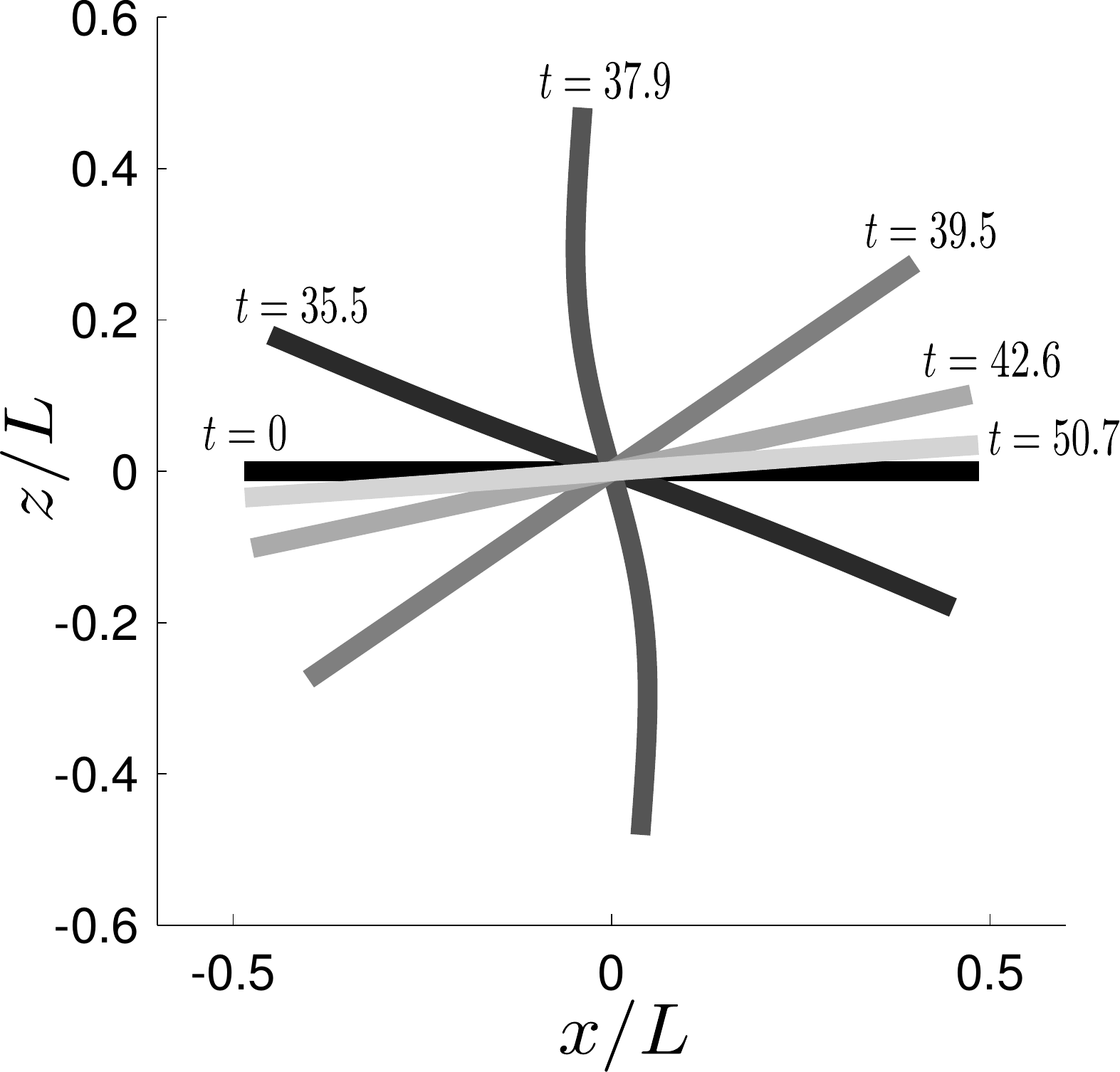}}
\subfloat[]{\label{fig:Snap_buckling} \includegraphics[height=6.5cm]{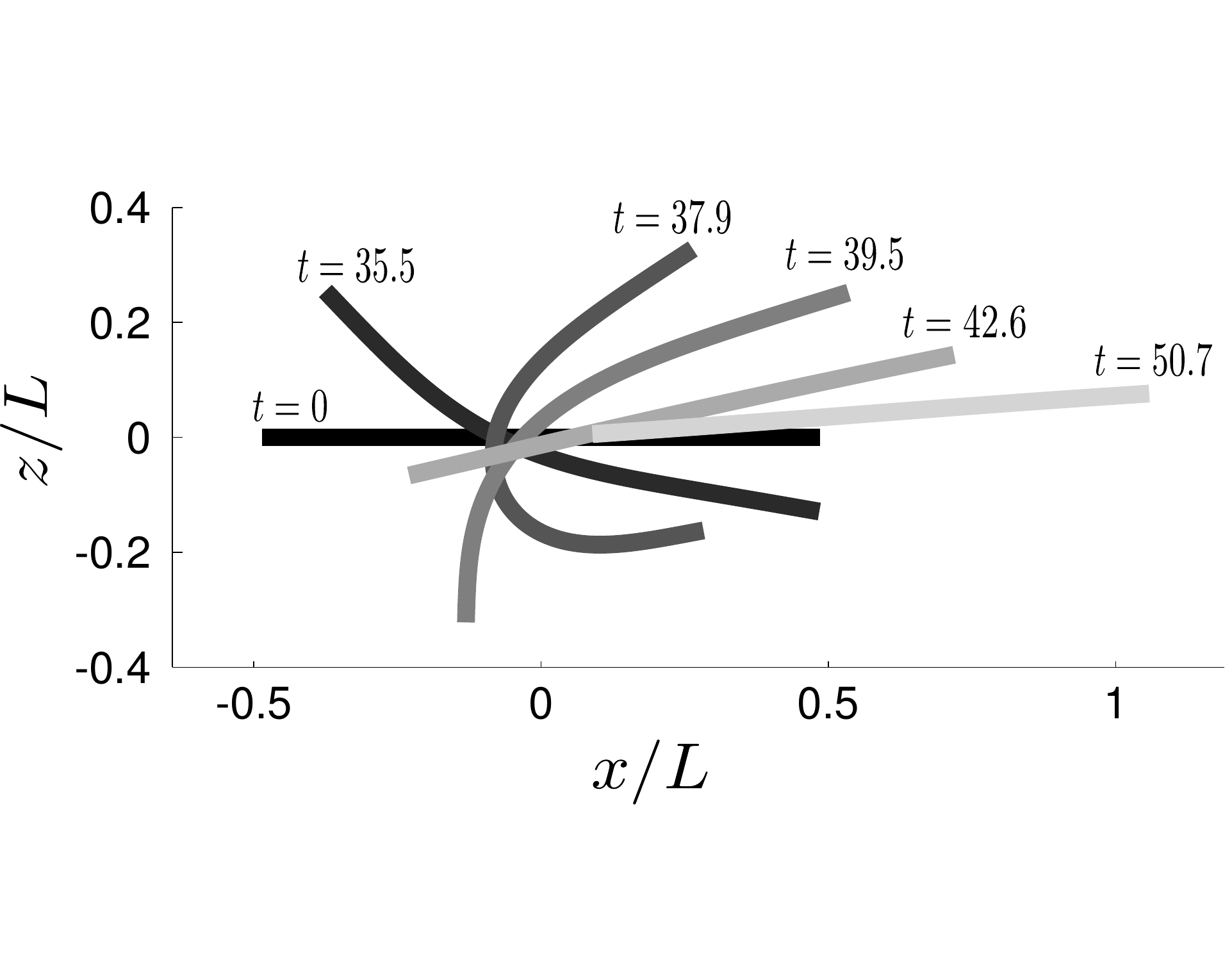}}
\par\end{centering}
\caption{
Orbit of a flexible filament in a shear flow  with BR = 0.04. Temporal evolution is shown in the plane of shear flow. 
a) Symmetric ``S-shape'' of a straight filament, $\kappa^{eq} = 0$. 
b) Buckling of a permanently deformed rod with an intrinsic curvature $\kappa^{eq} = 1/(100L)$.
}
\end{figure}

\cite{Forgacs1959a, Forgacs1959} analysed the motion of flexible threadlike particles in a shear flow depending on BR.
They observed important drifts from the Jeffery orbits and classified them into categories.
Yet, the goal of this section is not to carry out an in-depth study on these phenomena.
Instead, the objective is to show that our model can reproduce basic features characteristic of sheared flexible filaments. We analyze first the influence of intrinsic deformation on the motion.\\
If a fiber is straight at rest, it will symmetrically deform in a shear flow. 
When aligned with the compressive axes of the ambient rate of strain $\mathbf{E}^{\infty}$, the fiber adopts the ``S-shape'' observed in Fig. \ref{fig:Snap_buckling_no_int_curv}.
When aligned with the extensional axes, tensile forces turn the rod back to its equilibrium shape.
This symmetry is broken when the filament  is initially slightly deformed or has a permanent deformation at rest, i.e. a nonzero equilibrium curvature $\kappa^{eq}>0$ .
 An initial small perturbation of the shape of a straight filament aligned with flow can induce large deformations during the orbit. 
This phenomenon is known as the buckling instability whose onset and growth are quantified with BR \cite{Becker2001,Guglielmini2012}.
Fig. \ref{fig:Snap_buckling} illustrates the evolution of a flexible sheared filament with BR = 0.04 and a very small intrinsic deformation $\kappa^{eq} = 1/(100L)$.
The equilibrium dimensionless radius of curvature is $2R^{eq}/L=200$.
During the tumbling motion it decreases to a minimal value of $2R_{min}/L=0.26$. 
Buckling thus increases by 770 times the maximal fiber curvature.

\subsubsection{Maximal fiber curvature}
\cite{Salinas1981} measured the radius of curvature $R$ of sheared fiber for aspect ratios $r_p$ ranging from 283 to 680. 
They reported on the evolution of the minimal value $R_{min}$, i.e. the maximal curvature $\kappa_{max}$, with BR.
\cite{Schmid2000} used the Joint Model with prolate spheroids but no hydrodynamic interactions and compared  their results with \cite{Salinas1981}.
Both experimental results from \cite{Salinas1981} and simulations from \cite{Schmid2000} are accurately reproduced by the Gears Model.\\

Hydrodynamic interactions between fiber elements play an important role in the bending of flexible filaments \cite{Salinas1981, Lindstrom2007, Slowicka2012}.
As mentioned in Section \ref{section:Hydro_coupling} the use of spheres to build any arbitrary object is well suited to compute these hydrodynamic interactions.
However, modeling rigid slender bodies in a strong shear flow becomes costly when increasing the fiber aspect ratio.
First, the aspect ratio of a fiber made up of $N_b$ spheres is $r_{p}=N_b$.
Each time iteration requires the computation of $\mathcal{M}$ and $\mathcal{C}:\mathbf{E}^{\infty}$ and the inversion of a linear system \eqref{Lagrange_multiplier} corresponding to $N_c$ relations of constraints with $N_c\geq3\left(N_b-1\right)$.
Secondly, for a given shear rate $\dot{\gamma}$ and bending ratio BR, the Young's modulus increases as $r_{p}^{4}$. 
According to \eqref{eq:calc_dt}, the time step becomes very small for large $E$.
\cite{Schmid2000} partially avoided this issue by neglecting pairwise hydrodynamic interactions ($\mathcal{M}$ is diagonal), and by assembling prolate spheroids of aspect ratios $r_e\sim10$.\\

\begin{figure}
\begin{centering}
\subfloat[]{\label{fig:2Rmin_f_L} \includegraphics[height=6.3cm]{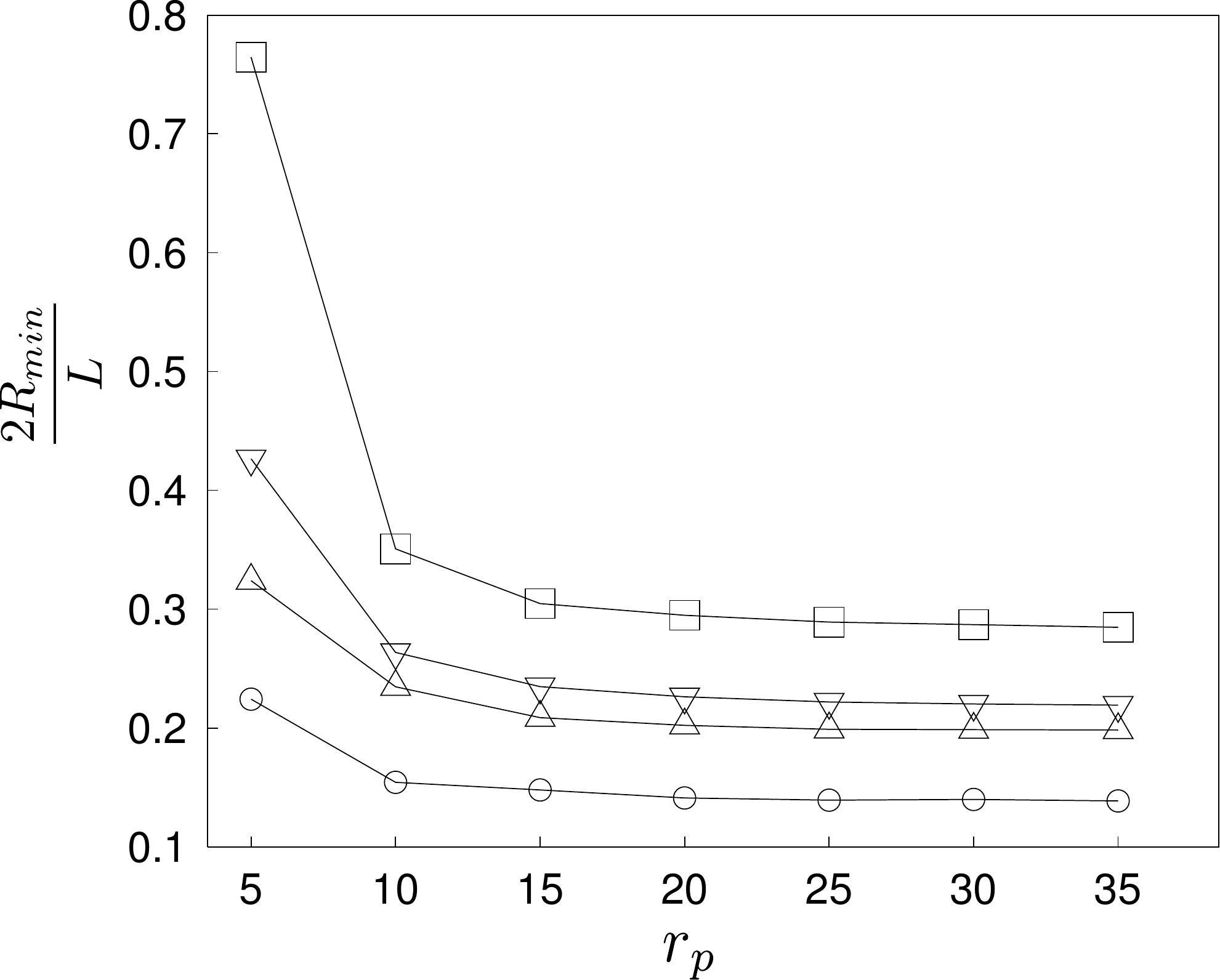}}
\subfloat[]{\label{fig:2Rmin_f_BR} \includegraphics[height=6.3cm]{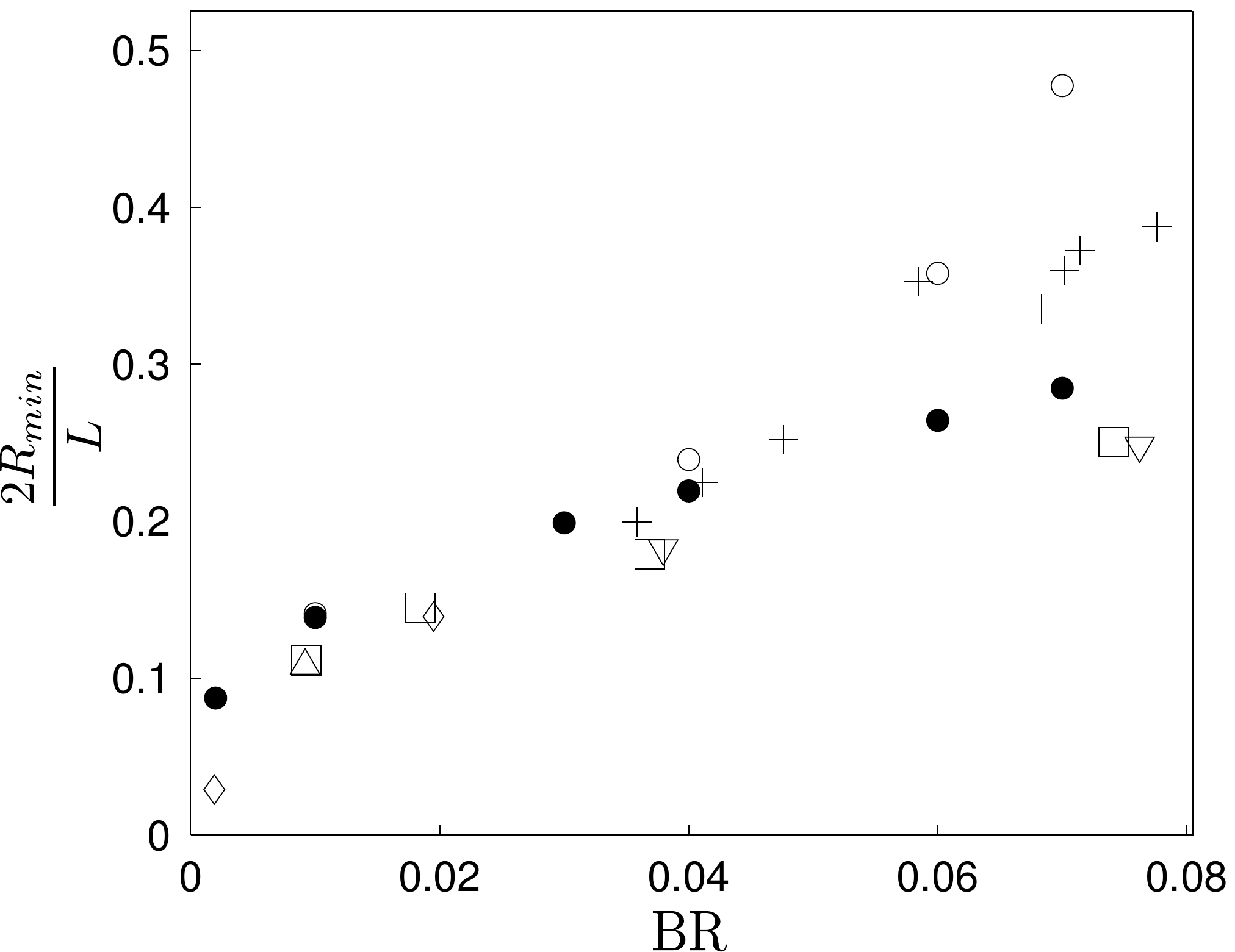}}
\par\end{centering}
\caption{
a) Minimal radius of curvature depending on fiber length for several bending ratios.
\symbol{\solid}{\ssquareb}{24}{1}{black}{black} : BR = 0.01, 
\symbol{\solid}{\circle}{24}{0}{black}{black} : BR = 0.03, 
\symbol{\solid}{\bigtriandown}{25}{0}{black}{black}: BR = 0.04 , 
\symbol{\solid}{\trianup}{25}{1}{black}{black}: BR = 0.07.
  b) Minimal radius of curvature along BR.
 \symbol{}{\circle}{-5}{0}{black}{black}: current simulations with aspect ratio $r_p = 35$ and intrinsic curvature $\kappa^{eq}=0$ ;
 \symbol{}{\blackcircle}{-5}{0}{black}{black}: current simulations with aspect ratio $r_p = 35$ and intrinsic curvature $\kappa^{eq}=1/(10L)$ ; 
 simulation results from \cite{Schmid2000} with $\kappa^{eq}=1/(10L)$: (\symbol{}{\losange}{-5}{0}{black}{black}: $r_p = 50$, 
							      \symbol{}{\trianup}{-5}{1}{black}{black}: $r_p = 100$, 
							      \symbol{}{\bigtriandown}{-5}{0}{black}{black}: $r_p = 150$, 
							      \symbol{}{\ssquareb}{-5}{1}{black}{black}: $r_p = 280$) ;
 \symbol{}{\plus}{-5}{0}{black}{black}: experimental measurements from \cite{Salinas1981}, $r_p = 283$. 
}
\end{figure}

Yet, it is shown in Fig. \ref{fig:2Rmin_f_L} that for a fixed BR, $2R_{min}/L$ converges asymptotically to a constant value with $r_p$. 
An asymptotic regime (relative variation less than $2\%$) is reached for $r_p\geq 25$. 
Choosing $r_p = 35$ thus enables a valid comparison with previous results.

Our simulation results compare well with the literature data (Fig. \ref{fig:2Rmin_f_BR}) and better match with to experiments than \cite{Schmid2000}. 
When $BR \geq 0.04$, the Gears Model clearly underestimates measurements for $\kappa^{eq} = 1/(10L)$ and overestimates
 them for  $\kappa^{eq} = 0$ . 
However, Salinas and Pittman \cite{Salinas1981} indicated that the error quantification on parameters and measurements is difficult to estimate as the fibers were hand-drawn. 
Notably, drawing accuracy decreases for large radii of curvature, which leads to the conclusion that the hereby observed  discrepancy might not be critical.
They did not report the value of permanent deformation $\kappa^{eq}$ for the fibers they designed, whereas, as evidenced 
by \cite{Forgacs1959}, it has a strong impact on $R_{min}$. 
A numerical study of this dependence should be conducted to compare with \cite{Forgacs1959}, Fig. 7.\\
\cite{Lindstrom2007} used the same approach as \cite{Schmid2000} with hydrodynamic interactions to repeat numerically  the experiments from \cite{Salinas1981} ; but their results, though reliable, were displayed such that direct comparison with previous  work is not possible.\\
To conclude, it should be noted that, in \cite{Schmid2000}, the aspect ratio does not affect $2R_{min}/L$ for a fixed BR, confirming the asymptotic behavior observed in Fig. \ref{fig:2Rmin_f_L}.\\

\subsection{Settling Fiber}

\begin{figure}
\begin{centering}
\subfloat[]{\label{fig:Metastable_B_10000} \includegraphics[height=6.5cm]{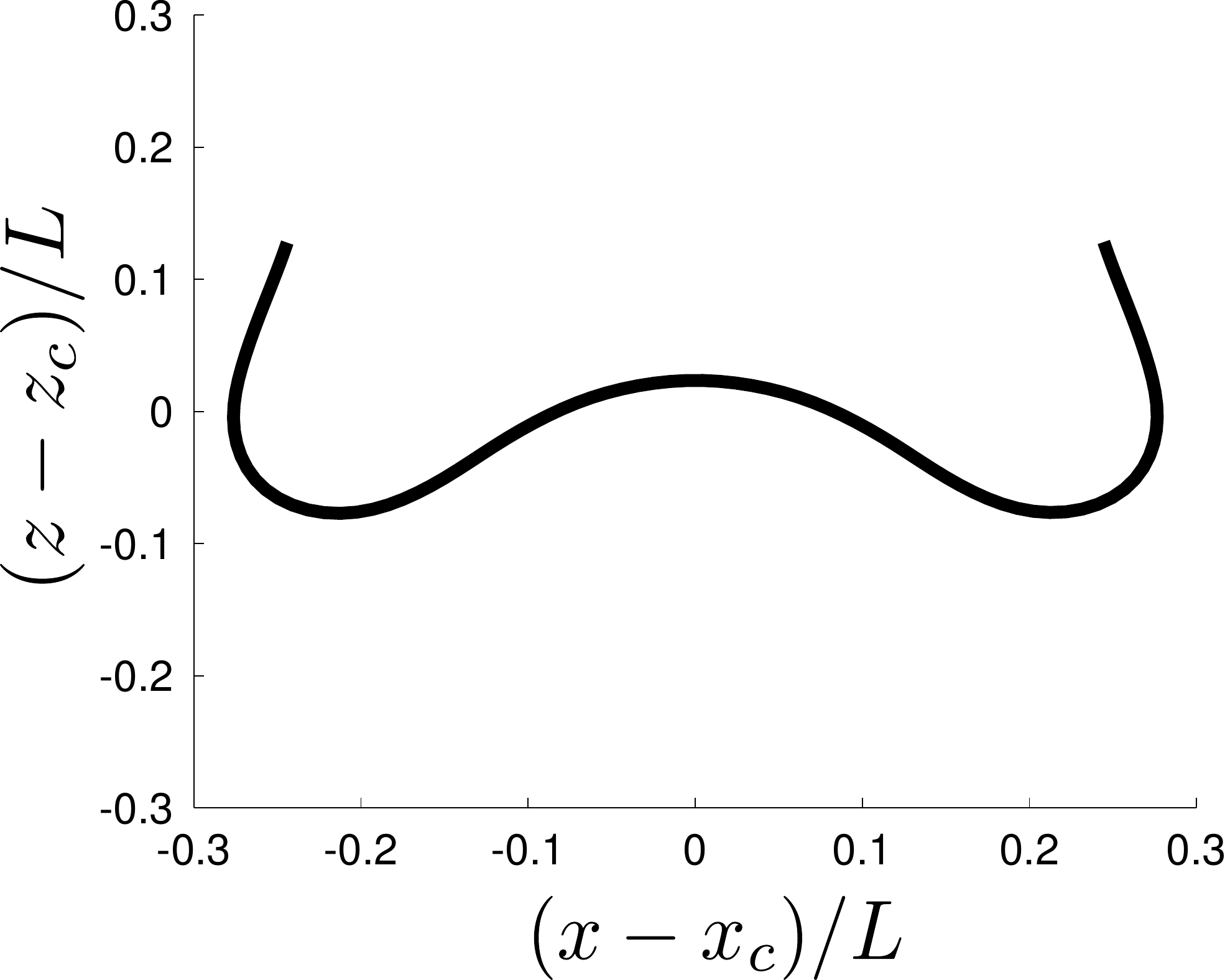}}
\subfloat[]{\label{fig:Stable_B_10000} \includegraphics[height=6.5cm]{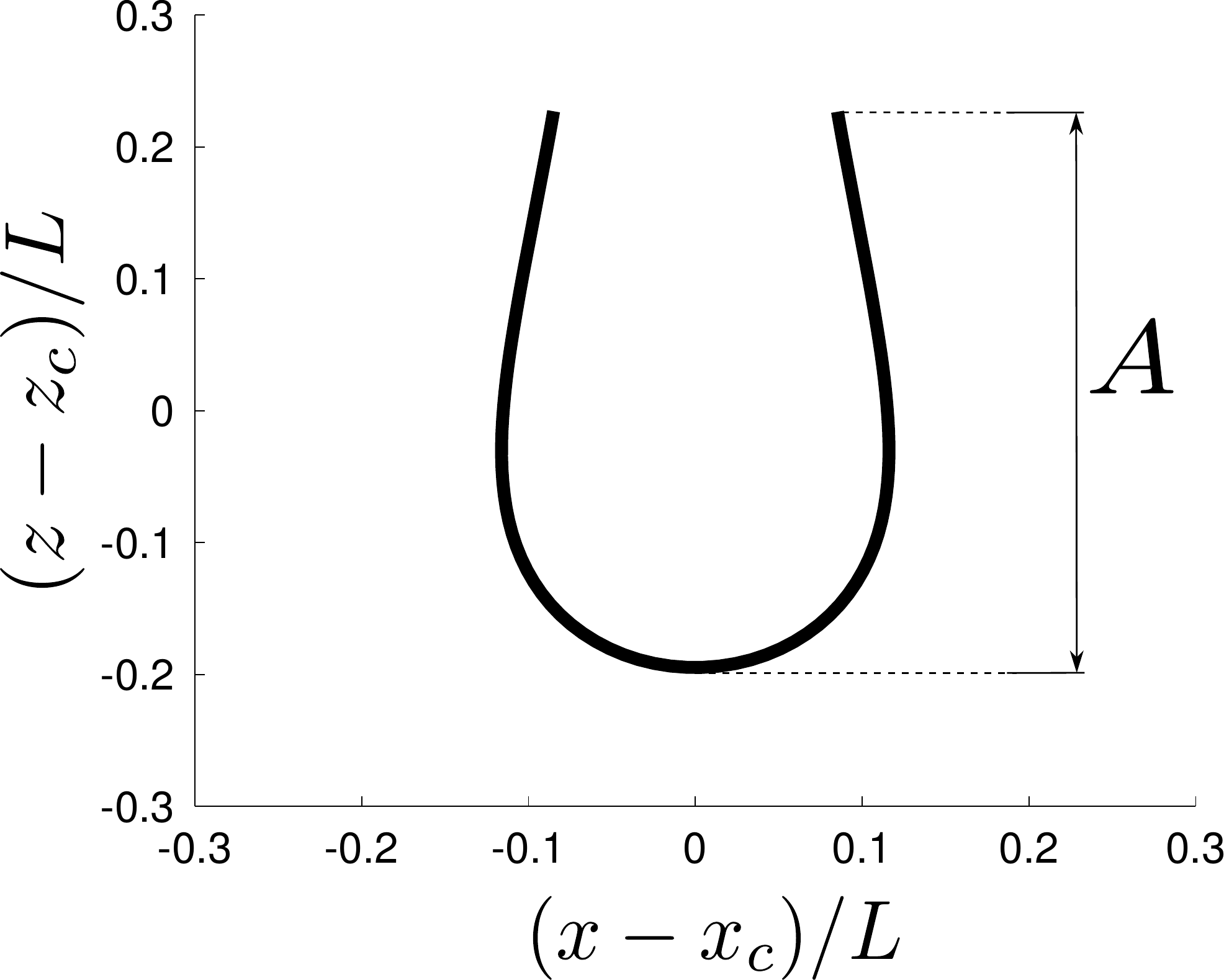}}
\par\end{centering}
\caption{ Shape of settling fiber for $B=10000$ in the frame moving with the center of mass $(x_c,z_c)$. a) Metastable ``W'' shape, $t=12 L/V_s$. b) Steady ``horseshoe'' shape at $t=53 L/V_s$. $V_s$ is the terminal settling velocity once steady state is reached.}
\end{figure}

Consider a fiber settling under constant gravity force $\mathbf{F}_{\perp} = F_{\perp}\mathbf{e}_{\perp}$ acting perpendicularly to its major axis. 
The dynamics of the system depends on three competing effects: the elastic stresses which tend to return the object to its equilibrium shape, the gravitational acceleration which uniformly translates the object and the hydrodynamic interactions which creates local drag along the filament. 
After a transient regime, the filament reaches steady state and settles at a constant velocity with a fixed shape (see Figs. \ref{fig:Metastable_B_10000} and \ref{fig:Stable_B_10000}). 
This steady state depends on the elasto-gravitational number 
\begin{equation}
B=F_{\perp}L/K^b. 
\end{equation}
\cite{CosentinoLagomarsino2005, Keaveny2008b} and \cite{Li2013} examined the contribution of each competing effect by measuring the normal deflection $A$, i.e. the distance between the uppermost and the lowermost point of the filament along the direction of the applied force (Fig. \ref{fig:Stable_B_10000}) ; and the normal friction coefficient $\gamma_{\perp}/\gamma^0_{\perp}$ as a function of $B$.
$\gamma^0_{\perp}$ is the normal friction coefficient of a rigid rod.
To compute hydrodynamic interactions \cite{CosentinoLagomarsino2005} used Stokeslet ; \cite{Keaveny2008b}, the Force Coupling Method (FCM) \cite{Yeo2010} ; \cite{Li2013}, Slender Body Theory.\\

Similar simulations were carried out with both the Joint Model described in Section \ref{Implementation_joint} and the Gears Model.
Fiber of length $L = 68a$ is made out of $N_b = 31$ beads with gap width $2\varepsilon_g = 0.2a$ for the Joint Model and $N_b = 34$ for the Gears Model.
To avoid both overlapping and numerical instabilities with the Joint Model, the following repulsive force coefficients were selected:
$d_0 = (a+\varepsilon_g)/4 $, $\delta_D = (a+\varepsilon_g)/5$, $C_1 = 0.01$ and $C_2=0.01$. No adjustable parameters are required for Gears Model.\\

\begin{figure}
\begin{centering}
\subfloat[]{\label{fig:Deflection} \includegraphics[height=8.6cm]{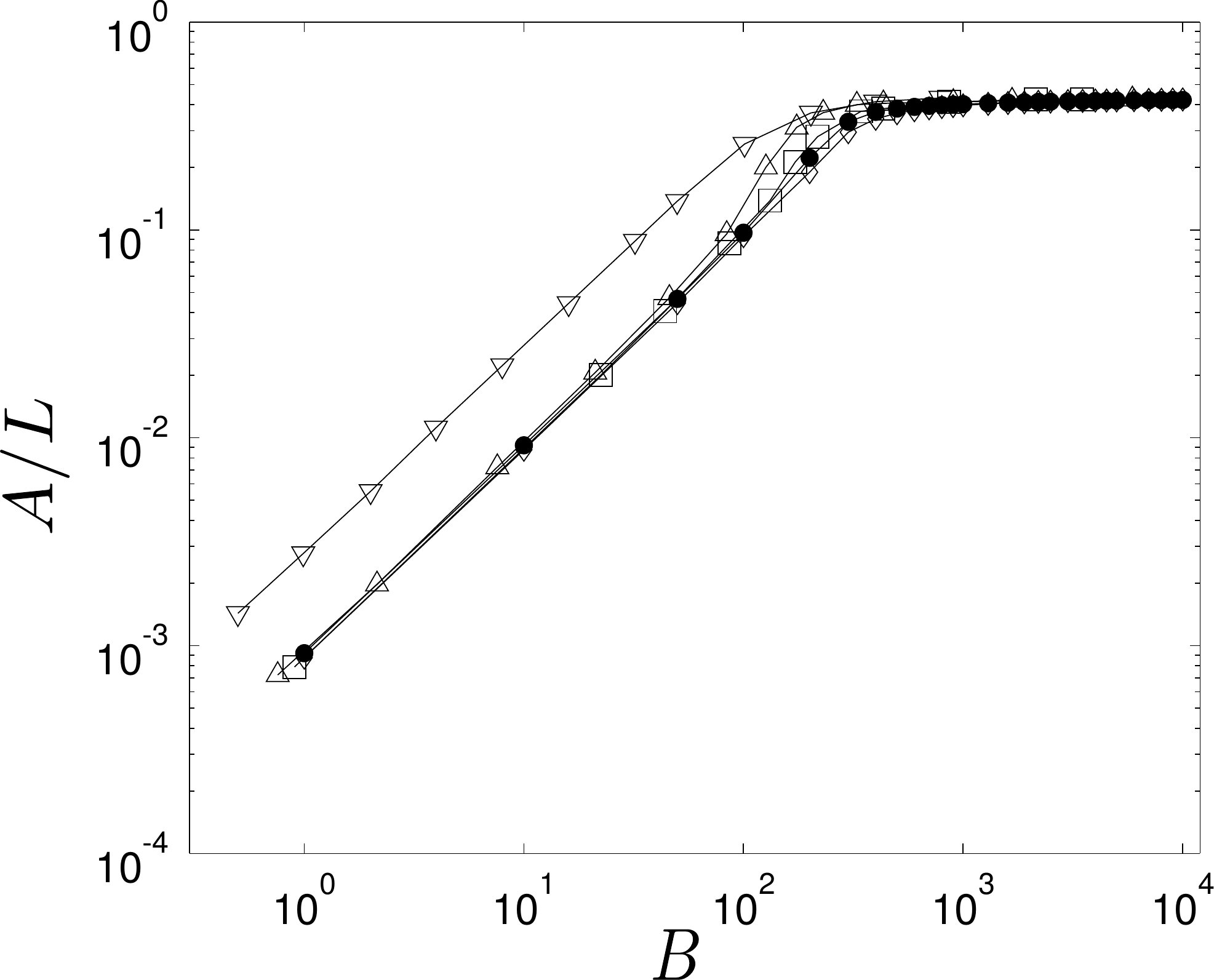}}
\\
\subfloat[]{\label{fig:Gamma} \includegraphics[height=8.5cm]{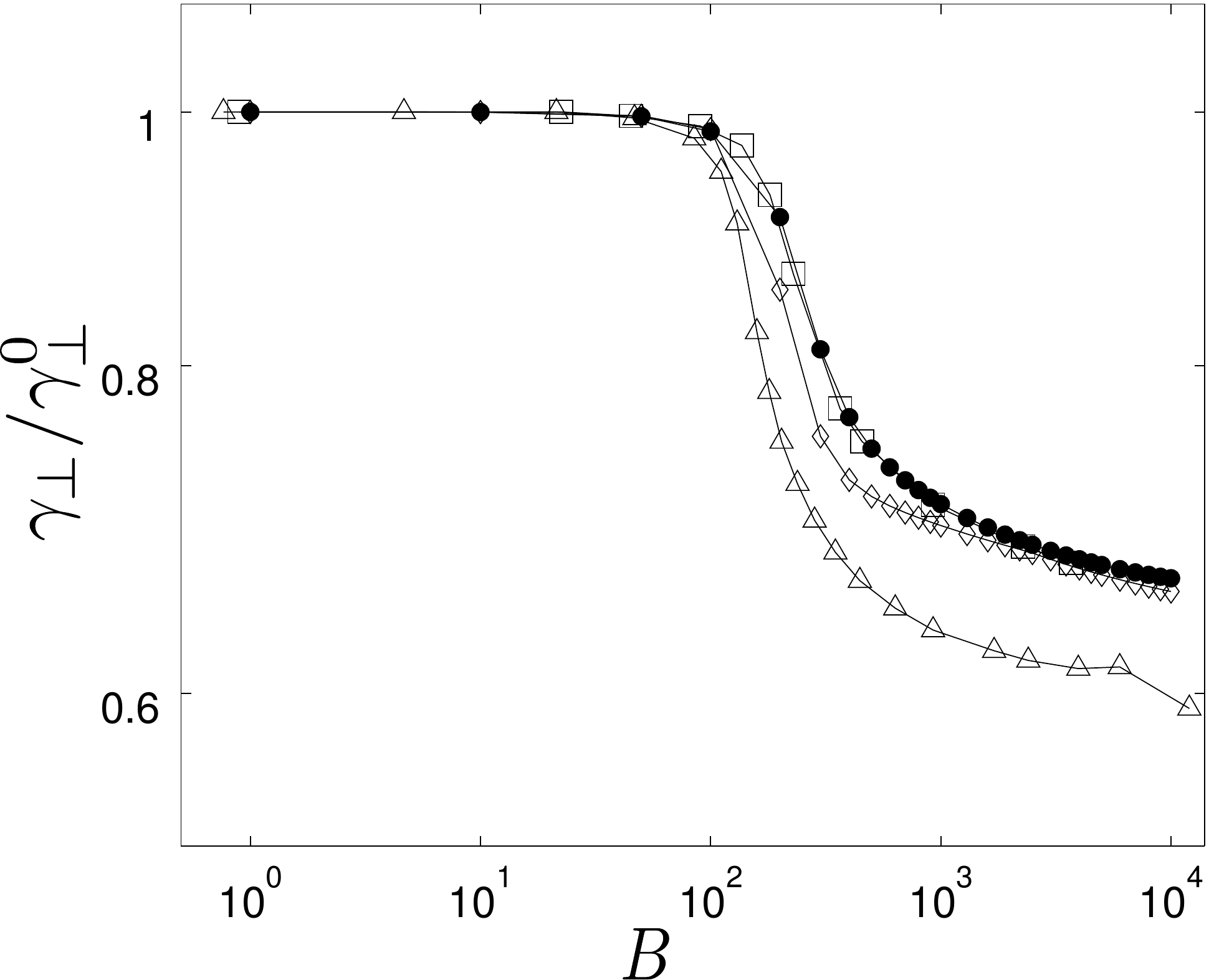}}
\par\end{centering}
\caption{
a) Scaled vertical deflection $A/L$ depending on $B$.
\symbol{\solid}{\blackcircle}{24}{0}{black}{black} : Gears Model, 
\symbol{\solid}{\losange}{24}{0}{black}{black} : Joint Model,
\symbol{\solid}{\ssquareb}{24}{1}{black}{black} : FCM results from \cite{Keaveny2008b}, 
\symbol{\solid}{\trianup}{24}{1}{black}{black} : Stokeslets results from \cite{CosentinoLagomarsino2005}, 
\symbol{\solid}{\bigtriandown}{24}{0}{black}{black} : Slender body theory results from \cite{Li2013}. 
b) Normal friction coefficient $vs.$ $B$.
\symbol{\solid}{\blackcircle}{24}{0}{black}{black} : Gears Model, 
\symbol{\solid}{\losange}{24}{0}{black}{black} : Joint Model,
\symbol{\solid}{\ssquareb}{24}{1}{black}{black} : FCM results from \cite{Keaveny2008b}, 
\symbol{\solid}{\trianup}{24}{1}{black}{black} : Stokeslets results from \cite{CosentinoLagomarsino2005}.
}
\label{fig:Comparisons_Settling}
\end{figure}

Fig. \ref{fig:Comparisons_Settling} shows that our simulations agree remarkably well with previous results except slight differences with \cite{Li2013} in the linear regime $B<100$. 
Using Slender Body Theory, \cite{Li2013} made the assumption of a spheroidal filament instead of a cylindrical one, with aspect ratio $r_p = 100$, i.e. three times larger than other simulations, whence such discrepancies.
The normal friction coefficient (Fig. \ref{fig:Gamma} ), resulting from hydrodynamic interactions, perfectly matches the value obtained by \cite{Keaveny2008b} with the Force Coupling Method.
The FCM is known to better describe multibody hydrodynamic interactions.
Such a result thus supports the use of the simple Rotne-Prager-Yamakawa tensor for this hydrodynamic system.\\

Differences between Gears and Joint Models implemented here are quantified by measuring the relative discrepancies on the vertical deflection $A$ 
\begin{equation}
\label{eq:dist_A}
 \epsilon_{A}=\dfrac{A_{G}-A_{J}}{A_{G}},
\end{equation}
and on the normal friction coefficient $\gamma_{\perp}/\gamma^0_{\perp}$
\begin{equation}
\label{eq:dist_gamma}
\epsilon_{\gamma^{\perp}}=\dfrac{\left(\gamma^{\perp}/\gamma_{0}^{\perp}\right)_{G}-\left(\gamma^{\perp}/\gamma_{0}^{\perp}\right)_{J}}{\left(\gamma^{\perp}/\gamma_{0}^{\perp}\right)_{G}}.
\end{equation}

Discrepancies between Joint and Gears models remain below $5\%$ except at the threshold of the non-linear regime ($B\approx100$) where $\epsilon_{A}$ reaches $15\%$ and $\epsilon_{\gamma^{\perp}} \approx 7.5\%$.\\

In accordance with \cite{CosentinoLagomarsino2005},  a metastable ``W'' shape is reached for $B>3000$  (Fig. \ref{fig:Metastable_B_10000}) until it converges to the stable ``horseshoe'' state (Fig. \ref{fig:Stable_B_10000}). 

%\begin{figure}
%\begin{centering}
%\includegraphics[height=7.5cm]{Relative_errors_settling_joint_gears.pdf}
%\par\end{centering}
%\caption{
%Discrepancies between Gears and Joint model implemented with the Euler-Lagrange formalism.
%\symbol{\solid}{\circle}{25}{0}{black}{black}: relative distance on the deflection $\epsilon_A$ \eqref{eq:dist_A}, 
%\symbol{\solid}{\ssquareb}{25}{0}{black}{black}: relative distance on the normal friction coefficient $\epsilon_{\gamma^{\perp}}$ \eqref{eq:dist_gamma}. 
%}
%\label{fig:Relative_dist}
%\end{figure}

\subsection{Actuated Filament}
\label{subsec:Actuated_Filament}
The goal of the following sections is to show that the model we proposed is not only valid for passive objects but also for active ones.
\emph{Elastohydrodynamics} also concern swimming at low Reynolds number \cite{Purcell1977}. 
Many type of micro-swimmers have been studied both from the experimental and theoretical point of view.
Among them two categories are distinguished: ciliates and flagellates.
Ciliates propel themselves by beating arrays of short hairs (cilia) on their surface in a synchronized way (\emph{Opalina, Volvox, Paramecia}).
Flagellates undulate and/or rotate their external appendage to push (pull) the fluid from their aft (fore) part (spermatozoa, \emph{Chlamydomonas Rheinardtii, Bacillus Subtilis, Eschericia Coli}).
Recent advances in nanotechnologies allows researchers to design artificial swimming micro-devices inspired by low Reynolds number fauna \cite{Dreyfus2005, Zhang2009, Keaveny2013}.\\
In that scope, the study of bending wave propagation along passive elastic filament has been investigated by \cite{Wiggins1998b}, \cite{Yu2006} and \cite{Coq2008, Coq2009}.\\

The experiment of \cite{Yu2006} consists in a flexible filament tethered and actuated at its base. 
The base angle was oscillated sinusoidally in plane with an amplitude $\alpha_0 = 0.435$rad and frequency $\zeta$.\\
Deformations along the tail result from the competing effects of bending and drag forces acting on it. 
A dimensionless quantity called  the Sperm number compares the contribution of viscous stresses to elastic response \cite{Wiggins1998}
\begin{equation}
\mbox{Sp}=L\left(\dfrac{\zeta\left(\gamma^{\perp}/L\right)}{K^{b}}\right)^{1/4}=\dfrac{L}{l_{\zeta}}.
\end{equation}
$\gamma^{\perp}$ is the normal friction coefficient. 
When using Resistive Force Theory, $\left(\gamma^{\perp}/L\right)$ is changed into a drag per unit length coefficient $\xi^{\perp}$. 
$l_{\zeta}$ can be seen as the length scale at which bending occurs.
Sp $\lesssim 1$ corresponds to a regime at which bending dominates over viscous friction: the whole filament oscillates rigidly in a reversible and symmetrical way.
Sp$\gg 1$ corresponds to a regime at which bending waves are immediately damped and the free end is motionless \cite{Wiggins1998}.\\

The experiment of \cite{Coq2008} is similar to  \cite{Yu2006} except that the actuation at the base is rotational.
Here, the filament was rotated at a frequency $\zeta$ forming a base angle $\alpha_0 = 0.262$rad with the rotation axis.\\

In both contributions, the resulting fiber deformations were measured and compared to Resistive Force Theory for several values of Sp. 
Simulations of such {\bf experiments \cite{Yu2006, Coq2008} } were performed with the Gears Model.\\

\subsubsection{Numerical setup and boundary conditions at the tethered base element}
Corresponding kinematic boundary conditions for BM are prescribed with the constraint formulation of the Euler-Lagrange formalism.\\

\paragraph{Planar actuation}
In the case of planar actuation  \cite{Yu2006}, we consider that the tethered, i.e. the first, bead is placed at the origin and has no degree of freedom 
\begin{equation}
\label{Constraint_bead_1}
\begin{cases}
\dot{\mathbf{r}}_{1}^{c}=\mathbf{0},\\
\boldsymbol{\omega}_{1}^{c}=\mathbf{0}.
\end{cases}
\end{equation}

Denote $\alpha_0$ the angle formed between $\mathbf{e}_x$ and $\mathbf{e}_{1,2}$.\\
The  trajectory of bead $2$ must follow 
\begin{equation}
 \mathbf{r}^c_{2}(t)=\left(\begin{array}{c}
2a\cos\left(\alpha_{0}\sin\left(\zeta t\right)\right)\\
0\\
2a\sin\left(\alpha_{0}\sin\left(\zeta t\right)\right)\end{array}\right).
\label{eq:2nd_bead_traj}
\end{equation}
The translational velocity of the second bead $ \dot{\mathbf{r}}_{2}(t) $ is thus constrained by taking the derivative of \eqref{eq:2nd_bead_traj}
\begin{equation}
\label{Constraint_bead_2}
 \dot{\mathbf{r}}^c_{2}(t)=\left(\begin{array}{c}
-2a\alpha_{0}\zeta\cos\left(\zeta t\right)\sin\left(\alpha_{0}\sin\left(\zeta t\right)\right)\\
0\\
2a\alpha_{0}\zeta\cos\left(\zeta t\right)\cos\left(\alpha_{0}\sin\left(\zeta t\right)\right)\end{array}\right).
\end{equation}
\\

\paragraph{Helical actuation}
In the case of helical beating \cite{Coq2008, Coq2009}, the anchor point of the filament 
is slightly off-centered with respect to the rotation axis $\mathbf{e}_x$ \cite{Coq2009}: $r(0) = \delta_0$ (cf. Fig. \ref{fig:Comparison_d_L}, Left inset).
\cite{Coq2009} measured a value $\delta_0 = 2mm$ with a filament length varying from $L = 2cm$ to $10cm$.
Here we take $\delta_0 =\tilde \delta_0\sin\alpha_{0}$ with $\tilde \delta_0=2.7 a $ and vary the filament length by changing the number of beads $N_b$ to match the experimental range $\delta_0/L = 0.1 \rightarrow 0.02$.
The position of bead $1$ must then follow 
\begin{equation}
 \mathbf{r}^c_{1}(t)=\left(\begin{array}{c}
\tilde \delta_0\cos\left(\alpha_{0}\sin\left(\zeta t\right)\right)\cos\left(\alpha_{0}\cos\left(\zeta t\right)\right)\\
\tilde \delta_0\cos\left(\alpha_{0}\sin\left(\zeta t\right)\right)\sin\left(\alpha_{0}\cos\left(\zeta t\right)\right)\\
\tilde \delta_0\sin\left(\alpha_{0}\sin\left(\zeta t\right)\right)\end{array}\right).
\label{eq:1st_bead_traj3D}
\end{equation}
The translational velocity of the first bead $ \dot{\mathbf{r}}_{1}(t) $ is thus constrained by taking the derivative of \eqref{eq:1st_bead_traj3D}
\begin{equation}
\label{Constraint_bead_1_3D}
 \dot{\mathbf{r}}^c_{1}(t)=\left(\begin{array}{c}
\tilde \delta_0\alpha_{0}\zeta\left[-\cos\left(\zeta t\right)\sin\left(\alpha_{0}\sin\left(\zeta t\right)\right)\cos\left(\alpha_{0}\cos\left(\zeta t\right)\right)\right.\\
             \qquad\quad\left. + \sin\left(\zeta t\right)\sin\left(\alpha_{0}\cos\left(\zeta t\right)\right)\cos\left(\alpha_{0}\sin\left(\zeta t\right)\right)\right]\\
\tilde \delta_0\alpha_{0}\zeta\left[-\cos\left(\zeta t\right)\sin\left(\alpha_{0}\sin\left(\zeta t\right)\right)\sin\left(\alpha_{0}\cos(\zeta t)\right)\right.\\
             \qquad\quad\,\left. - \sin\left(\zeta t\right)\cos\left(\alpha_{0}\cos\left(\zeta t\right)\right)\cos\left(\alpha_{0}\sin\left(\zeta t\right)\right)\right]\\
\tilde \delta_0\alpha_{0}\zeta\cos\left(\zeta t\right)\cos\left(\alpha_{0}\sin\left(\zeta t\right)\right)\end{array}\right).
\end{equation}
And the rotational velocity is set to zero $\boldsymbol{\omega}_1 = \mathbf{0}$.\\
The velocity of the second bead $\dot{\mathbf{r}}^c_{2}(t)$ is prescribed in synchrony with bead 1:
\begin{equation}
\label{Constraint_bead_2_3D}
 \dot{\mathbf{r}}^c_{2}(t)=\left(\begin{array}{c}
(\tilde \delta_0+2a)\alpha_{0}\zeta\left[-\cos\left(\zeta t\right)\sin\left(\alpha_{0}\sin\left(\zeta t\right)\right)\cos\left(\alpha_{0}\cos\left(\zeta t\right)\right)\right.\\
             \qquad\quad\left. + \sin\left(\zeta t\right)\sin\left(\alpha_{0}\cos\left(\zeta t\right)\right)\cos\left(\alpha_{0}\sin\left(\zeta t\right)\right)\right]\\
(\tilde \delta_0+2a)\alpha_{0}\zeta\left[-\cos\left(\zeta t\right)\sin\left(\alpha_{0}\sin\left(\zeta t\right)\right)\sin\left(\alpha_{0}\cos\left(\zeta t\right)\right)\right.\\
             \qquad\quad\,\left. - \sin\left(\zeta t\right)\cos\left(\alpha_{0}\cos\left(\zeta t\right)\right)\cos\left(\alpha_{0}\sin\left(\zeta t\right)\right)\right]\\
(\tilde \delta_0+2a)\alpha_{0}\zeta\cos\left(\zeta t\right)\cos\left(\alpha_{0}\sin\left(\zeta t\right)\right)\end{array}\right).
\end{equation}
The rotational velocity $\boldsymbol{\omega}_2$ is consistently constrained by the no-slip condition.
The three-dimensional curvature $\boldsymbol{\kappa}$ is discretized with \eqref{eq:curv_3D}. \\

In both cases, imposing actuation at the base of the filament therefore requires the addition of three vectorial kinematic constraints, \eqref{Constraint_bead_1} and \eqref{Constraint_bead_2}, to the no-slip conditions: $N_c = 3(N_b-1) + 3\times3$.
The additional Jacobian matrix $\mathbf{J}^{act}$  writes
\begin{equation}
\mathbf{J}^{act}=\left(\begin{array}{ccccccc}
\mathbf{I}_{3} & \mathbf{0}_{3} & \mathbf{0}_{3} & \mathbf{0}_{3} & \cdots & \mathbf{0}_{3} & \mathbf{0}_{3}\\
\mathbf{0}_{3} & \mathbf{I}_{3} & \mathbf{0}_{3} & \mathbf{0}_{3} & \cdots & \mathbf{0}_{3} & \mathbf{0}_{3}\\
\mathbf{0}_{3} & \mathbf{0}_{3} & \mathbf{I}_{3} & \mathbf{0}_{3} & \cdots & \mathbf{0}_{3} & \mathbf{0}_{3}\end{array}\right).
\end{equation}

The corresponding right-hand side $\mathbf{B}^{act}$ contains the imposed velocities 
\begin{equation}
\mathbf{B}^{act}=\left(\begin{array}{c}
\mathbf{0}\\
\mathbf{0}\\
-\dot{\mathbf{r}}_{2}^{c}\end{array}\right)
\end{equation}
for planar beating, and
\begin{equation}
\mathbf{B}^{act}=\left(\begin{array}{c}
-\dot{\mathbf{r}}_{1}^{c}\\
\mathbf{0}\\
-\dot{\mathbf{r}}_{2}^{c}\end{array}\right)
\end{equation} 
for helical beating.

$\mathbf{J}^{act}$ and  $\mathbf{B}^{act}$ are simply appended to $\mathcal{J}$ and $\mathbf{B}$ respectively; corresponding forces and torques $\mathfrak{F}_c$ are computed as explained before in Section \ref{section:Euler Lagrange}.

\subsubsection{Comparison with experiments and theory}

The dynamics of the system can be described by balancing elastic stresses (flexion and tension) with viscous drag.
Subsequent coupled non-linear equations can be linearized with the approximation of small deflections or solved with an adaptive integration scheme \cite{Camalet2000, Lauga2007, Gadelha2010a}.

\begin{figure}
\begin{centering}
\includegraphics[height=17cm]{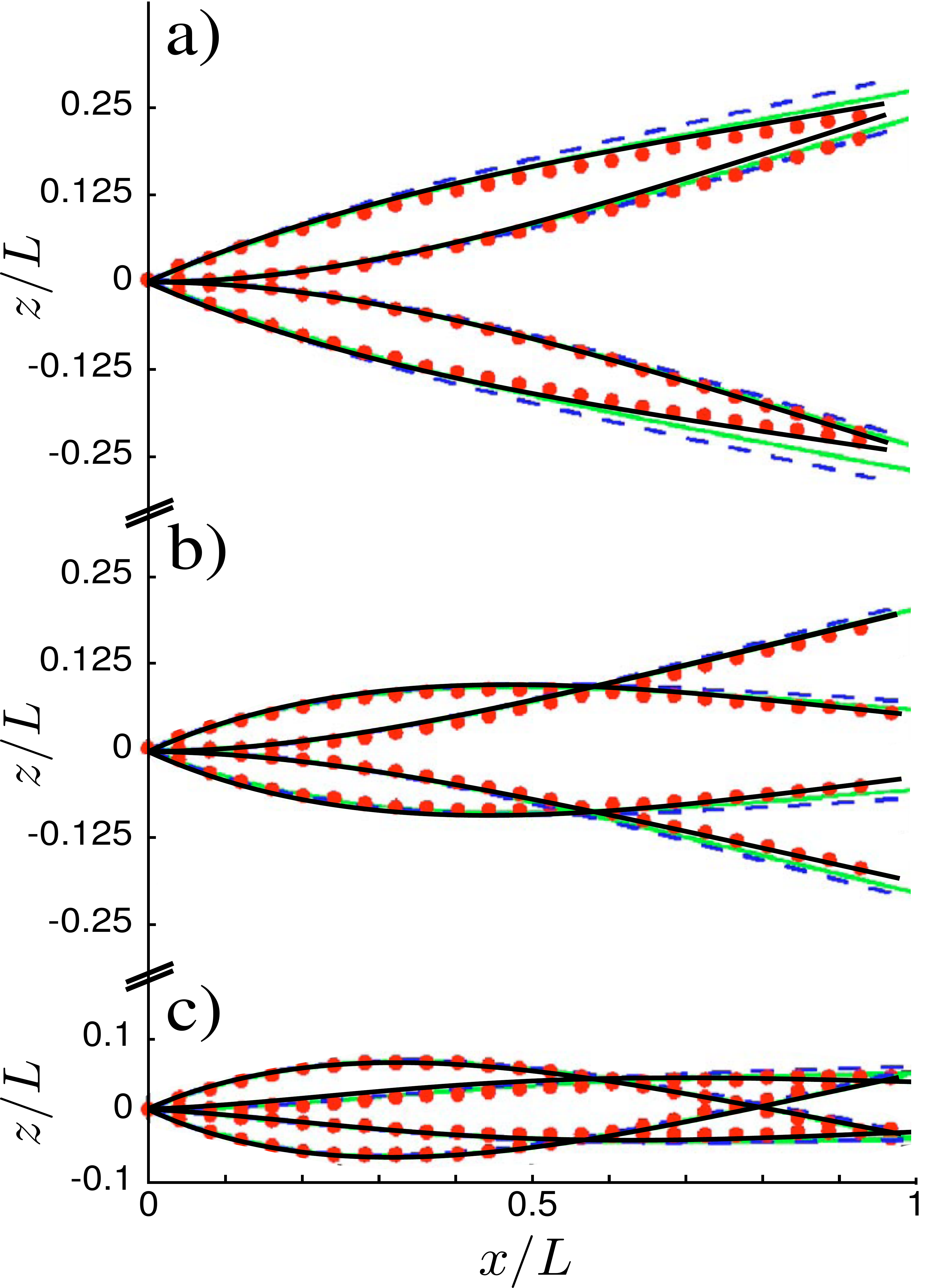}
\par\end{centering}
\caption{Comparison with experiments and numerical results from \cite{Yu2006}. Gears Model results are superimposed on the original Fig. 3 of \cite{Yu2006}.  
Snapshots are shown for four equally spaced intervals during the cycle for one tail with $\alpha_0 = 0.435$rad. 
\symbol{}{\blackcircle}{-5}{0}{black}{red}: experiment,
\symbol{\solid}{}{8}{0}{green}{black}: linear theory,
\symbol{\dashed}{}{8}{0}{blue}{black}: non-linear theory,
\symbol{\solid}{}{8}{0}{black}{black}: Gears Model,
a) $\zeta = 0.5$ rad$.s^{-1}$, Sp$=1.73$. 
b) $\zeta = 1.31$ rad$.s^{-1}$, Sp$=2.2$. 
c) $\zeta = 5.24$ rad$.s^{-1}$, Sp$=3.11$.}
\label{fig:Comparison_actuated}
\end{figure}

\paragraph{Planar actuation}
\cite{Yu2006} considered both linear and non-linear theories and included the effect of a sidewall by using the corrected RFT coefficients of \cite{Mestre1975}.\\
Simulations are in good agreement with experiments, linear and non-linear theories for Sp $= 1.73, \, 2.2, \,$ and $3.11$ (Fig. \ref{fig:Comparison_actuated}).
Even though sidewall effects were neglected here, the Gears Model provides a good description of non-linear dynamics of an actuated filament in Stokes flow.

\begin{figure}
\begin{centering}
\includegraphics[height=10cm]{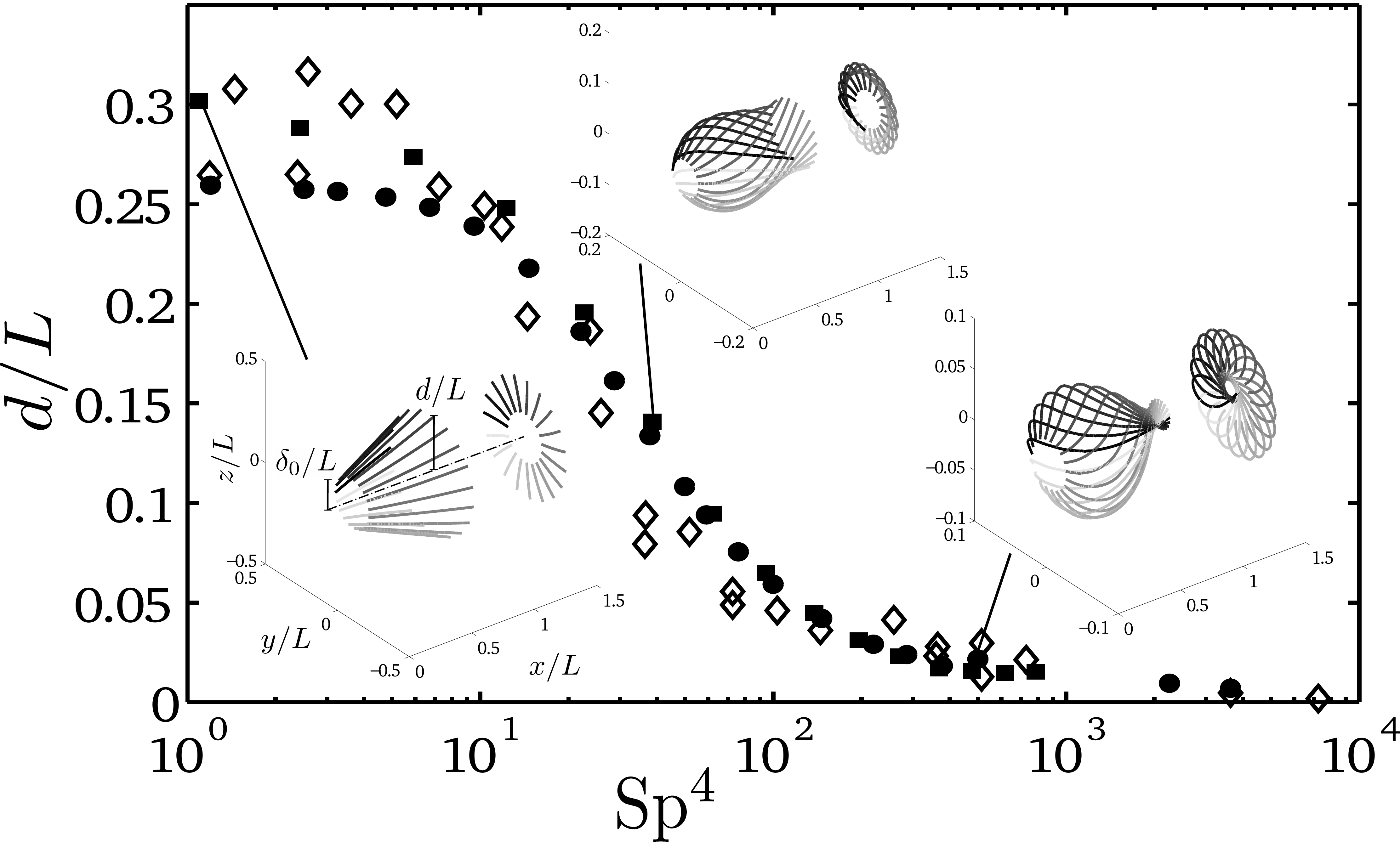}
\par\end{centering}
\caption{  Comparison with experiments from \cite{Coq2008}.
(Insets) Evolution of the filament shape with Sp$^4$. Snapshots are shown for twenty equally spaced intervals during one period at steady state. 
Gray level fades as time progresses. 
Left inset: $\delta_0/L$ is the distance of the tethered bead to the rotation axis,  $d/L$ is the distance of the free end to the rotation axis. 
(Main figure) Distance of the rod free end to the rotation axis normalized by the filament length $d/L$. 
\symbol{}{\losange}{-5}{0}{black}{black}: experiment,
\symbol{\blackcircle}{}{8}{0}{black}{black}: Gears Model with no anchoring distance $\delta_0/L = 0$,
\symbol{\blackssquare}{}{8}{0}{black}{black}: Gears Model with $\delta_0/L = 0.1 \rightarrow 0.02$ as in \cite{Coq2009}. 
}
\label{fig:Comparison_d_L}
\end{figure}

\paragraph{Helical actuation}
Once steady state was reached, \cite{Coq2008} measured the distance of the tip of the rotated filament to the rotation axis $d = r(L)$ (cf. Fig. \ref{fig:Comparison_d_L}, Left inset). 
Figure \ref{fig:Comparison_d_L} compares their measures with our numerical results. 
Insets show the evolution of the filament shape with Sp. 
The agreement is quite good. 
Numerical simulations slightly overestimate $d$ for $30 < \text{Sp}^4 < 90$. 
This may be due  to the lack of information to reproduce experimental conditions and/or
to measurement errors.
As stated in \cite{Coq2009}, taking the anchoring distance $\delta_0$ into account is important to match the low Sperm number configurations where $\delta_0/L$ is non-negligible and the filament is stiff.
If the anchoring point was aligned with the rotation axis ($\delta_0=0$), the distance to the axis of the rod free end would be $d/L = \sin \alpha_0 = 0.259$ for small Sp, as shown on Fig. \ref{fig:Comparison_d_L}.

\subsection{Planar swimming Nematode}
\label{Planar_Swimming_Alone}

\begin{figure}
\begin{centering}
\subfloat[]{ \label{fig:Motion_nematode}  \raisebox{2.3cm}{\includegraphics[height=3.5cm]{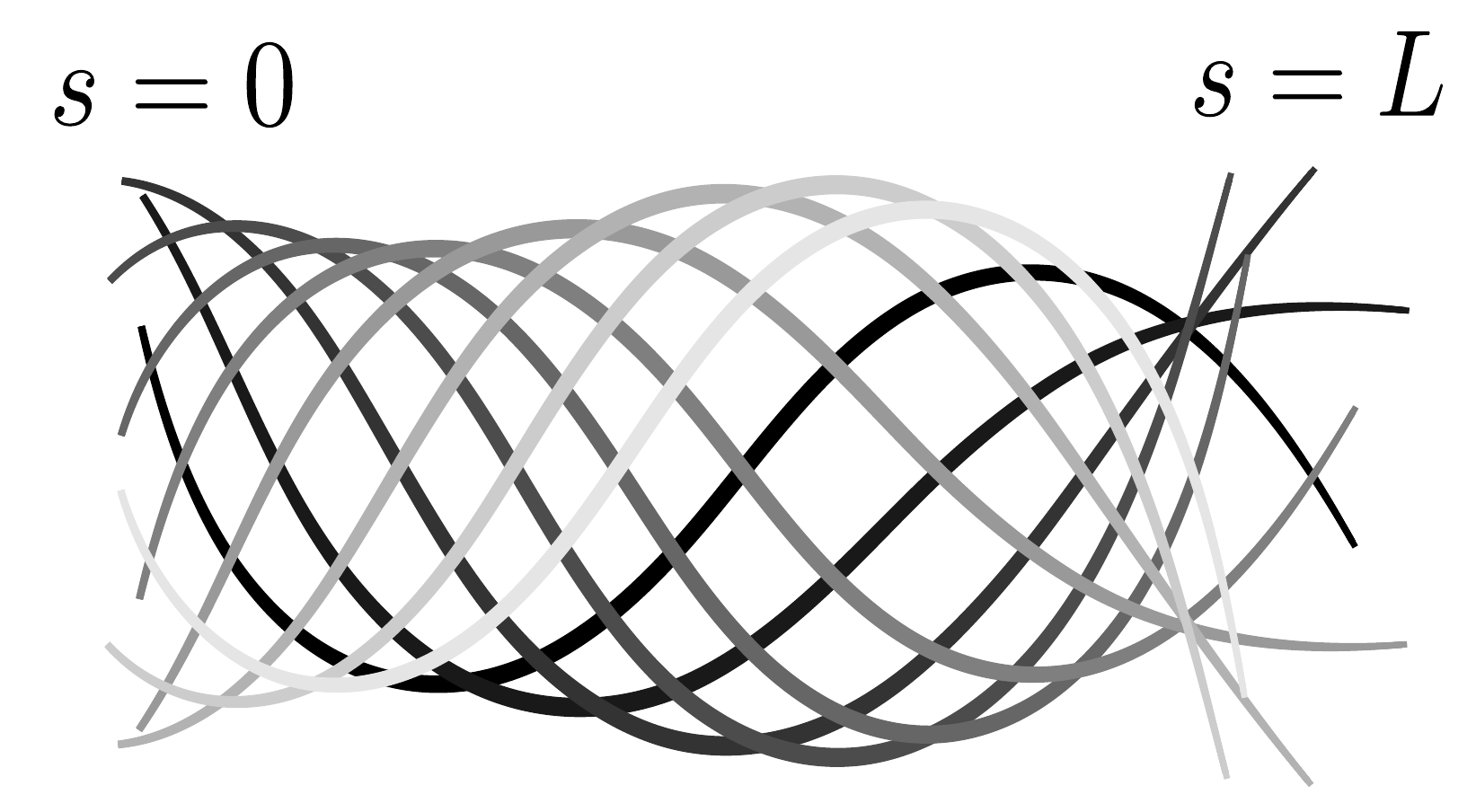}}}
\hspace{0.5cm}
\subfloat[]{ \label{fig:Two_swimmers} \includegraphics[height=7.5cm]{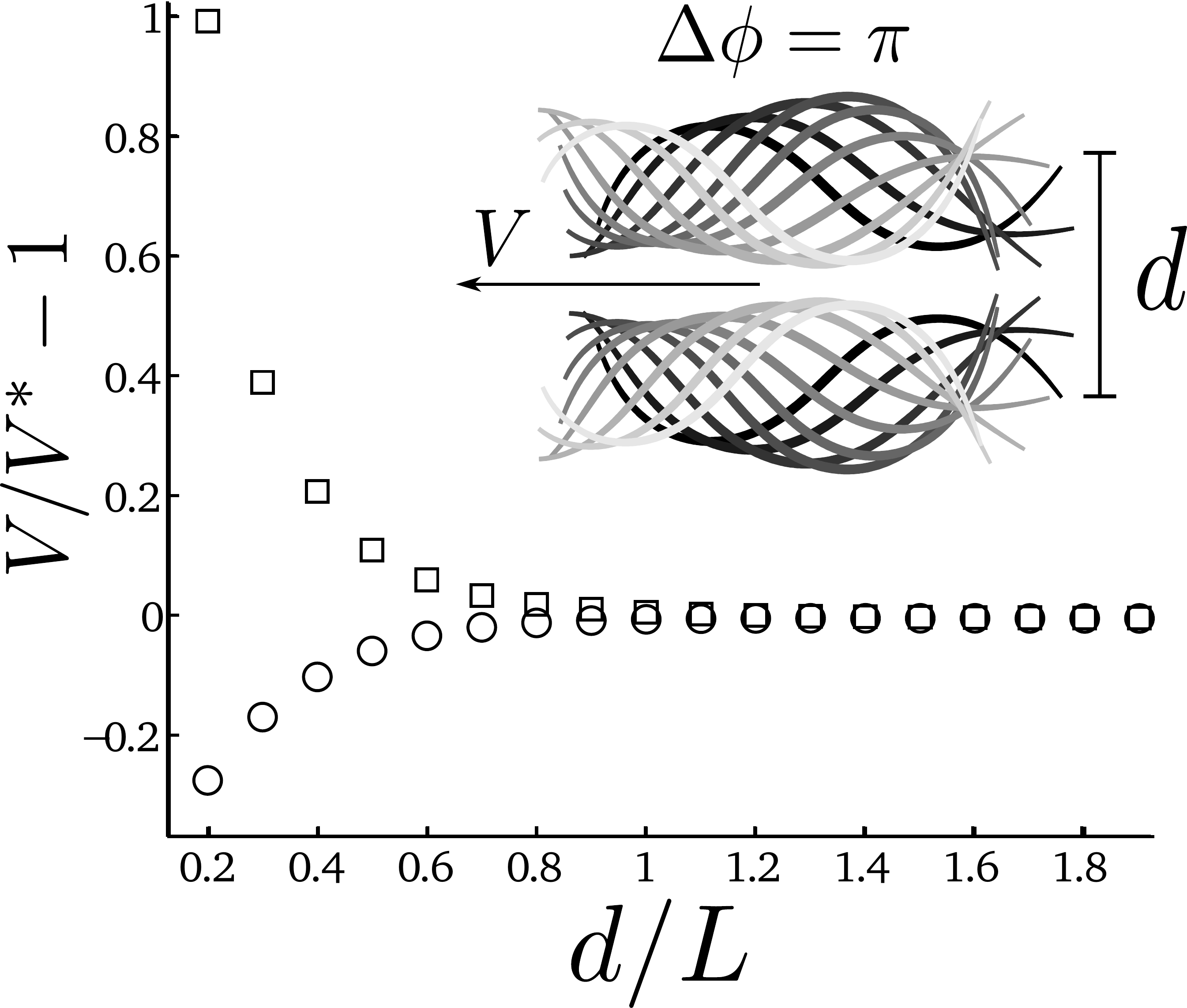}}
\par\end{centering}
\caption{a) Simulated wave motion of a swimming model \emph{C. Elegans}. 
The nematode swims leftward and gray level fades as time progresses. 
Motion is shown in a frame moving with the microswimmer center of mass.
 b) (Inset) Two \emph{C. Elegans} beating in the same plane at a distance $d$ in opposite phase ($\Delta \phi = \pi$). 
Nematodes swim leftward and gray level fades as time progresses. 
(Main figure) Swimming speed of the center of mass of the system $V$ normalized by the isolated swimming speed of \emph{C. Elegans} $V^*$.
\symbol{}{\circle}{-5}{0}{black}{black}: in-phase motion $(\Delta \phi = 0)$ ; 
\symbol{}{\ssquareb}{-5}{0}{black}{black}: antiphase motion $(\Delta \phi = \pi)$.} 
\end{figure}

Locomotion of the nematode \emph{Caenorhabditis Elegans} is addressed here as its dynamics and modeling are well documented \cite{Majmudar2012, Bilbao2013}.
\emph{C. Elegans} swims by propagating a contraction wave  along its body length, from the fore to the aft (Fig. \ref{fig:Motion_nematode}).
modeling such an active filament in the framework of BM just requires the addition of an oscillating driving torque $\gamma^D(s,t)$ to mimic the internal muscular contractions.
To do so, \cite{Majmudar2012} used the preferred curvature model.
In this model, the driving torque results from a deviation in the centerline curvature from 
\begin{equation}
 \kappa^{D}(s,t)=-\kappa_{0}^{D}(s)\sin\left(ks-2\pi ft\right),
 \label{eq:Sp_keav}
\end{equation}
where $\kappa_{0}^{D}(s)$ is prescribed to reproduce higher curvature near the head:
\begin{equation}
 \kappa_{0}^{D}\left(s\right)=\begin{cases}
				K_{0}, & s\leq0.5L\\
				2K_{0}\left(L-s\right)/L, & s>0.5L.\end{cases}
\end{equation}

The amplitude $K_0$, wave number $k$  and the associated Sperm number 
 
\begin{equation} \text{Sp} = L\left(f\mu/K^b \right)^{1/4}
\end{equation}
were tuned to reproduce the measured curvature wave of the free-swimming nematode.
They obtained the following set of numerical values: $K_0 = 8.25/L$, $k = 1.5\pi/L$ and  $\text{Sp}^* = 22.6^{1/4}$ .
The quantity of interest to compare with experiments is the distance the nematode travels per stroke $V/(fL)$.
$K^b$ is assumed to be constant along $s$ and is deduced from the other parameters.
As for \eqref{eq:Bending_torque_bead}, the torque applied on bead $i$ results from the difference in active bending moments across neighbouring links
\begin{equation}
 \boldsymbol{\gamma}_{i}^{D}(t)=\mathbf{m}^D\left(s_{i+1},t\right)-\mathbf{m}^D\left(s_{i-1},t\right),
\end{equation}
with $\mathbf{m}^D\left(s_{i},t\right) = K^b\kappa^D(s_i,t)\mathbf{b}(s_i)$. $\boldsymbol{\gamma}^D$ is added to $\mathfrak{F}_a$ at step 3 of the algorithm in Section \ref{sec:Algo}.

To match the aspect ratio of \emph{C. Elegans}, $r_p = 16$, \cite{Majmudar2012} put $N_b = 15$ beads together separated by gaps of width $2\epsilon_g =0.2a$.
Here we assemble $N_b=16$ beads, avoiding the use of gaps, and employ the same target-curvature wave and numerical coefficient values.\\

The net translational velocity $V^* = V/(fL) = 0.0662$ obtained with our model matches remarkably well with the numerical results $V/(fL) = 0.0664$ \cite{Majmudar2012} and experimental measurements $V/(fL) \approx 0.07$ \cite{Bilbao2013}.

\subsection{Cooperative swimming}
 One of the configurations explored in \cite{Llopis2013} has been chosen as a test case for the interactions between
 in-phase or out-of phase swimmers. Two identical, coplanar \emph{C. Elegans} swim in the same direction with a phase difference $\Delta \phi$ which is introduced in the target curvature, and thus in the driving torque, of the second swimmer 
\begin{equation}
 \kappa^{D,2}(s,t)=-\kappa_{0}^{D}(s)\sin\left(ks-2\pi ft + \Delta \phi\right).
\end{equation}
 The initial shape of the swimmers is taken from their steady state. We define $d$ as the distance between their center of mass at initial time (see inset of Fig. \ref{fig:Two_swimmers}).
 Similarly to \cite{Llopis2013} Fig. 3, our results (Fig. \ref{fig:Two_swimmers}) show that antiphase beating enhance the propulsion, whereas in-phase swimming slows the system as swimmers get closer. 
 Even though the model swimmer here is different, the quantitative agreement with \cite{Llopis2013} is strikingly good. 
 Numerical work by \cite{Fauci1990} also revealed that the average swimming speed of infinite sheets in finite Reynolds number flow is maximized when they beat in opposite phase.
 The conclusion that closer swimmers do not necessarely swim faster than 
individual ones has also been reported in \cite{Immler2007}. 
They measured a decrease in the swimming speed of $25\%$ for groups of house mouse sperm, as obtained on Fig. \ref{fig:Two_swimmers} for $d/L=0.2$.

\subsection{Spiral swimming}

 Many of the flagellate microorganisms such as spermatozoa, bacteria or artificial micro-devices use spiral swimming to propel through viscous fluid. 
Propulsion with rotating rigid or flexible filaments has been thoroughly investigated in the past years (\cite{Cortez2005,Manghi2006, Qian2008, Keaveny2008, Coq2009, Hsu2009, Keaveny2013}).
In this section we illustrate the versatility of the proposed model by investigating the effect of the Sperm number and the eccentricity of the swimming gait on the swimming speed of \emph{C. Elegans}.

\subsubsection{Numerical configuration}
The target curvature of \emph{C. Elegans} $\kappa^{D}$ remains unchanged except that it is now directed along two components which are orthogonal to the helix axis. 
A phase difference $\Delta \phi = \pi/2$ is introduced between these two components. 
The resulting driving moment writes:
\begin{equation}
 \mathbf{m}^{D}\left(s_{i},t\right)=\alpha K^{b}\kappa^{D}(s_{i},t)\mathbf{e}_{\bot}+\beta K^{b}\kappa^{D}(s_{i},t,\Delta\phi=\pi/2)\mathbf{e}_{b}.
\end{equation}
$\{\mathbf{e}_{\Vert},\mathbf{e}_{\bot},\mathbf{e}_{b}\}$ are body fixed orthonomal vectors. 
$\mathbf{e}_{\Vert}$ is directed along the axis of the helix, $\mathbf{e}_{\bot}$ is a perpendicular vector and $\mathbf{e}_{b}$ is the binormal vector completing the basis (Fig. \ref{fig:Vsw_Helical} inset).
The magnitude of the curvature wave along $\mathbf{e}_{\bot}$ (resp. $\mathbf{e}_{b}$) is weighted by a coefficient $\alpha$ (resp. $\beta$). 
%The ratio $\beta/\alpha$ is linked to the eccentricity  $e$ of the swimming gait in the plane $\{\mathbf{e}_{\bot},\mathbf{e}_{b}\}$ by the relation $e = \sqrt{1-(\beta/\alpha)^2}$.
The trajectory of a body element in the plane $\{\mathbf{e}_{\bot},\mathbf{e}_{b}\}$  describes an ellipse whose eccentricity depends on the value of the ratio $\beta/\alpha$.
When $\beta/\alpha=0$ the driving torque is two-dimensional and identical to the one used in Section \ref{Planar_Swimming_Alone}. 
When $\beta/\alpha=1$ the magnitude of the driving torque is equal in both direction, the swimming gait describes a circle in the plane $\{\mathbf{e}_{\bot},\mathbf{e}_{b}\}$ (see Fig. \ref{fig:Vsw_Helical} inset).
For the sake of simplicity, here we take $\{\mathbf{e}_{\Vert},\mathbf{e}_{\bot},\mathbf{e}_{b}\} = \{\mathbf{e}_{x},\mathbf{e}_{y},\mathbf{e}_{z}\}$.
As in \ref{subsec:Actuated_Filament}, the curvature is evaluated with \eqref{eq:curv_3D}.
In the following, $\alpha = 1$ and only $\beta$ is varied in the range $[0;1]$.

\subsubsection{Results}
\begin{figure}
\begin{centering}
\includegraphics[height=8.5cm]{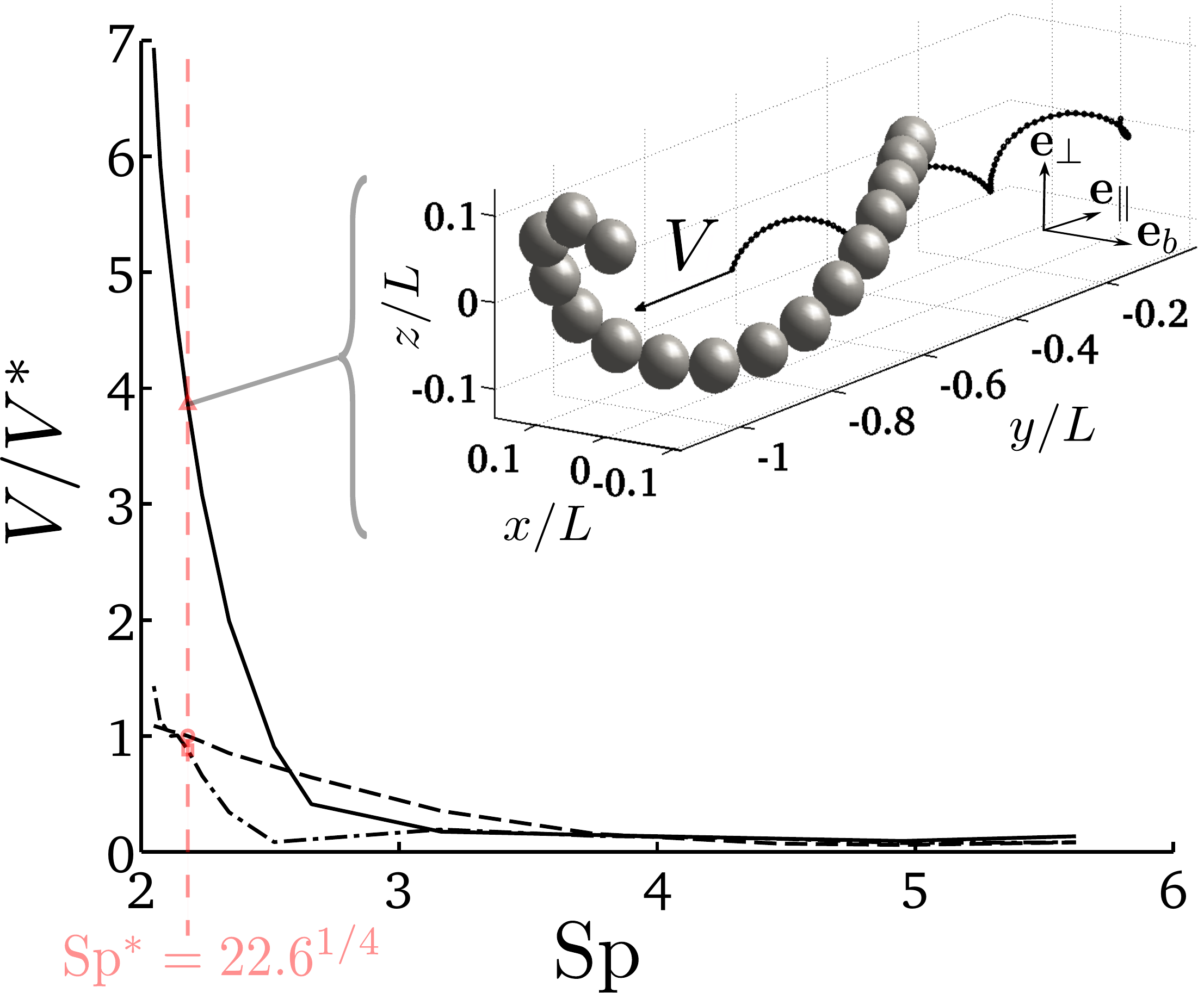}
\par\end{centering}
\caption{  Helical swimming of \emph{C. Elegans}. (Inset) Snapshot for Sp$^* = 22.6^{1/4}$ and $\beta/\alpha = 1$. 
\symbol{\solid}{\blackcircle}{22}{0}{black}{black}: trajectory of the center of mass.
(Main figure) Swimming speed of the center of mass $V$ normalized by the planar swimming speed of \emph{C. Elegans} $V^*$.
\symbol{\solid}{}{9}{0}{black}{black}: $\beta/\alpha = 1$,
\symbol{\dashdot}{}{10}{0}{black}{black}: $\beta/\alpha = 0.5$,
\symbol{\dashed}{}{11}{0}{black}{black}: $\beta/\alpha = 0$ (planar motion),
} 
\label{fig:Vsw_Helical}
\end{figure}

Figure \ref{fig:Vsw_Helical} compares the planar swimming speed of \emph{C. Elegans} $V^*$, with its ``helical'' version $V$, depending on the Sperm number defined in Section \ref{Planar_Swimming_Alone} \eqref{eq:Sp_keav} and on the ratio $\beta/\alpha$.
The Sperm number Sp lies in the range $[17^{1/4};1000^{1/4}]=[2.03;5.62]$. 
The lower bound is dictated by the stability of the helical swimming. 
When $\text{Sp}<2.03$, the imposed curvature reaches a value such that the swimmer experiences a change in shape which is not helical.
This sudden change in shape breaks any periodical motion and makes irrevelant the measurement of a net translational motion.
Such limitation is only linked to the choice of the numerical coefficients of the target curvature model.

For the characteristic value Sp$^*=22.6^{1/4}$ chosen by \cite{Majmudar2012}, the purely helical motion provides a swimming speed four times faster than planar beating. 
Even though the model swimmer is different here, this result qualitatively agrees with the observation of \cite{Keaveny2008} for which spiral swimming was faster than planar beating.
Beyond a critical value Sp $\approx2.6$, planar beating is faster. 
For $\beta/\alpha = 0.5$ the swimming speed is always smaller than for planar beating except when Sp $<2.15$. 
This last observation is not intuitive. 
A more extensive study on the effect of the eccentricity of the swimming gait on the swimming speed would be of interest.

\section{Conclusions}

We have provided a simple %, Lagrange multiplier oriented, 
general theoretical framework for kinematic constraints to be used in three-dimensional BM.
%, which, to our knowledge, has not been previously given.
This framework permits to handle versatile and complex kinematic constraints between flexible assembly of spheres, and/or more complex non deformable objects
 %coupled by their hydrodynamic interactions 
 at low Reynolds numbers. 
 Using Stokes linearity, this formulation requires, at each time step, the inversion of a $O(N_c \times N_c)$ linear system 
 %built upon the product of mobility matrix to constraints matrix 
 for an assembly having $N_c$ constraints.  
Constraints are  exactly matched (up to machine precision)  and their evaluation is insensitive to time-step. 
Furthermore, since the formulation explicitly handles mobility matrices,  it can be used with any  approximation for hydrodynamics interactions, from free drain (no HI) to full Stokesian Dynamics. 
The proposed framework also implicitly incorporates the  generic influence of external flows  on kinematic constraints, as opposed to previous BM formulation  which necessitates some adjustments to the
ambient flow in most cases.

We also  propose 
%in this contribution 
a simple Gears Model to describe flexible objects, and we show that such model  successfully predicts the fiber dynamics in an external flow, its response to an external mechanical forcing and the motion of internally driven swimmers. 
Quantitative agreement with previous works is obtained for both slender objects (fibers, actuated filaments) and non-slender swimmers (\emph{C. Elegans}), allowing its use in a wide variety of contexts. 
The Gears Model is easy to implement and fulfills several important improvements over previous BM :
\begin{itemize}
\item There is no limitation on the fiber curvature, since Gears Model does not need any repulsive force  nor gap width to be defined.
\item Gears Model is more generic than previous ones, since there is no need for numerical parameter to be tuned.  
%It is worth mentioning that parameter  adjustments are needed in previous BM which depend on the  external flow.
\item When compared with Lagrange multiplier formulation of Joint Model, Gears Model is also much more stable by two orders of magnitude in time-step, a drastic improvement which offers nice prospects for the modeling of complex flexible assemblies.
\end{itemize}

Finally it should be noted that even if we only consider  simple  collections of spheres, any  complex assembly can be easily treated within a similar framework, which
provide interesting prospects in the future modeling of 
complex microorganims,  membranes or  cytoskeleton micro-mechanics.

\section*{Acknowledgements}

The authors would like to acknowledge support from the ANR project MOTIMO: Seminal Motility Imaging and Modeling. 
We are also thankful to Pr Pierre Degond, Yuan-nan Young and Eric E. Keaveny for fruitful discussions.

\appendix

\section{Correspondence between 
 ${\bf M}$ and  ${\bf  \mathcal M} $}
\label{appendix-sec1}
Matrix ${\bf  \mathcal M} $ defined  in \eqref{mobilite_generalized} results from the rearrangement of the well-known mobility matrix $\mathbf{M}$.
This operation is necessary in order to combine constraint equation  \eqref{non-holonomous_constraint} and mobility relation \eqref{mobilite} to obtain the constraint forces $\mathfrak{F}_c$.\\

Matrix $\mathbf{M}$ relates the collection of velocities $\mathbf{V}=\left(\mathbf{v}_1,...,\mathbf{v}_{N_b}\right)$ and rotations $\boldsymbol{\Omega}=\left(\boldsymbol{\omega}_1,...,\boldsymbol{\omega}_{N_b}\right)$ to the collection of forces  $\mathbf{F}=\left(\mathbf{f}_1,...,\mathbf{f}_{N_b}\right)$ and torques $\mathbf{\Gamma}=\left(\boldsymbol{\gamma}_1,...,\boldsymbol{\gamma}_{N_b}\right)$
\begin{equation}
 \left(\begin{array}{c}
\mathbf{V}\\
\boldsymbol{\Omega}\end{array}\right)=\left(\begin{array}{cc}
\mathbf{M}^{VF} & \mathbf{M}^{VT}\\
\mathbf{M}^{\Omega F} & \mathbf{M}^{\Omega T}\end{array}\right)\left(\begin{array}{c}
\mathbf{F}\\
\boldsymbol{\Gamma}\end{array}\right),
\label{Mobility_relation}
\end{equation}

where $\mathbf{M}^{VF}$ is the $3N_b\times 3N_b$ matrix relating all the bead velocities to the forces applied to their center of mass
\begin{equation}
 \mathbf{M}^{VF}=\left(\begin{array}{ccc}
\mathbf{M}_{11}^{VF} & \ldots & \mathbf{M}_{1N_{b}}^{VF}\\
\vdots & \ddots & \vdots\\
\mathbf{M}_{N_{b}1}^{VF} & \ldots & \mathbf{M}_{N_{b}N_{b}}^{VF}\end{array}\right).
\end{equation}

\eqref{Mobility_relation} is not consistent with the structure of the generalized velocity $\dot{\mathbf{Q}} = \left(\mathbf{v}_1,\boldsymbol{\omega}_1,...,\mathbf{v}_{N_b},\boldsymbol{\omega}_{N_b}\right)$ and force $\mathfrak{F} = \left(\mathbf{f}_1,\boldsymbol{\gamma}_1,...,\mathbf{f}_{N_b},\boldsymbol{\gamma}_{N_b}\right)$ vectors.
Thus we rearrange $\mathbf{M}$ into $\mathcal{M}$ such that
\begin{equation}
 \mathcal{M}_{ii}=\left(\begin{array}{cc}
\mathbf{M}_{ii}^{VF} & \mathbf{M}_{ii}^{VT}\\
\\\mathbf{M}_{ii}^{\Omega F} & \mathbf{M}_{ii}^{\Omega T}\end{array}\right),
\end{equation}

to obtain a mobility equation suited for the Euler-Lagrange formalism
\begin{equation}
 \left(\begin{array}{c}
\dot{\mathbf{q}}_{1}\\
\vdots\\
\dot{\mathbf{q}}_{N_b}\\\end{array}\right)=\left(\begin{array}{ccc}
\mathcal{M}_{11} & \ldots & \mathcal{M}_{1N_{b}}\\
\vdots & \ddots & \vdots\\
\mathcal{M}_{N_{b}1} & \ldots & \mathcal{M}_{N_{b}N_{b}}\end{array}\right)\left(\begin{array}{c}
\mathfrak{f}_{1}\\
\vdots\\
\mathfrak{f}_{N_b}\end{array}\right).
\label{mobility_generalized_details}
\end{equation}

\eqref{mobility_generalized_details} is strictly equivalent to \eqref{mobilite_generalized}.

\section{Asymptotic limit of force and moment balance on the Gears Model}
\label{appendix-sec2}

In this appendix we show that slender body formulation for elastic fibers, when applied to Gears Model is consistent with the discrete formulation of force and moments balance  (\ref{total_force_bead_i}) and (\ref{total_moment}) in the asymptotic limit of small beads.

The force balance equation for a beam is \cite{Landau1975}
\begin{equation}
\label{force_balance_ll}
 \dfrac{\partial\mathbf{n}^{is}}{\partial s}+\mathbf{f}=\mathbf{0},
\end{equation}
$\mathbf{n}^{is}(s)$ is the resultant internal stress on a cross-section $S(s)$ at arclength position $s$ along the centerline
\begin{equation}
\label{N_def}
\mathbf{n}^{is}(s)=\int_{S(s)}  \boldsymbol{\sigma} \cdot {\bf t} \,\, dS,
\end{equation}
for which the tangent vector to neutral fiber centerline is also the unit normal vector to cross-section $S(s)$. $\mathbf{f}$ is the force per unit length which contains any supplementary contribution to the internal elastic response of the material (e.g. hydrodynamic force per unit length). The moment  balance  reads \cite{Landau1975}
\begin{equation}
\label{moment_balance_ll}
 \dfrac{\partial\mathbf{m}^{is}}{\partial s}+\mathbf{t}\times\mathbf{n}^{is}+\boldsymbol{\tau}=\mathbf{0},
\end{equation}
where ${\mathbf m}^{is}(s)$ is the moment of the flexion and torsion stresses on the cross-section which are related to the local deformation of Frenet-Serret  coordinates 
%\begin{equation}
%\label{M_def}
%\mathbf{M}^{is}(s)=\int_{S(s)}  {\bf x} \times {\bf \sigma} \cdot {\bf t} \,\, dS,
%\end{equation}
and $\boldsymbol{\tau}$  is the torque 
per unit length resulting from supplementary contributions to the internal elastic response.

Let us consider the curvilinear  integral of  (\ref{force_balance_ll}) over each bead $i$, following the centerline of the skeleton joining  the contact point $c_{i-1}$ between bead $i-1$ and $i$ and the bead center $\mathbf{r}_i$, as well as the  bead center and the contact point $c_{i}$ between bead $i$ and $i+1$ (see Figure \ref{fig:Sketch_gears_model}). The curvilinear arclength $s$ thus varies from $2ai$ to $2a(i+1)$ within bead $i$. 

Following the centerline, the integral of the internal stress contribution to \eqref{force_balance_ll} reads
\begin{equation}
\label{N_int}
 \int_{2ai}^{2a(i+1)} \dfrac{\partial\mathbf{n}^{is}}{\partial s} ds=\mathbf{n}^{is}(2a(i+1))-\mathbf{n}^{is}(2ai).
\end{equation}

In the limit of pointwise contacts, the normal stress produced by  contact
forces at contact point $c_{i-1}$ located at point ${\bf x}_{c_{i-1}}$ reads
\begin{equation}
\label{boundary_condition}
 \boldsymbol{\sigma} \cdot {\bf t}={\bf f }_{c_{i-1}} \delta_S ({\bf x}-{\bf x}_{c_{i-1}}),
\end{equation}

where $\delta_S$ stands for the surface Dirac distribution at the bead surface. Consequently the moment distribution associated with a Dirac contact forces  applied at ${\bf x}_{c_{i-1}}$ is
\begin{equation}
\label{boundary_condition_moment}
{\bf m }_{c_{i-1}} \delta_S ({\bf x}-{\bf x}_{c_{i-1}})=a {\bf t}_{i-1} \times {\bf f }_{c_{i-1}}\delta_S ({\bf x}-{\bf x}_{c_{i-1}}).
\end{equation}

Since the area of the cross-section $S(s)$ normal to the centerline tends to the bead surface itself as $s \rightarrow 2ai$ or $ s \rightarrow 2a(i+1)$, one can find that ${\bf n}^{is}(2 a i) \rightarrow{\bf f }_{c_{i-1}}$ and  ${\bf n}^{is}(2 a (i+1)) \rightarrow {\bf f }_{c_{i}}$ as $a \rightarrow 0$ using (\ref{N_int}), (\ref{boundary_condition}) and (\ref{N_def}). 
Hence, the finite size integral of (\ref{force_balance_ll}), fulfills the following limit as the bead radius tends to zero 
\begin{equation}
\label{force_balance_asympt}
{\bf f }_{c_{i-1}} -{\bf f }_{c_{i}}+ {\bf f }=0,
\end{equation}
which is consistent with the force used in (\ref{total_force_bead_i}).

The second term of the moment balance equation (\ref{moment_balance_ll}) from contact point $c_{i-1}$ to bead center $\mathbf{r}_i$ is
\begin{equation}
\label{tvecNi}
 \int_{2ai}^{2a(i+1/2)}\mathbf{t}\times\mathbf{n}^{is} \, ds = \mathbf{t}_{i-1}\times \int_{2ai}^{2a(i+1/2)}\mathbf{n}^{is} \, ds=\mathbf{t}_{i-1}\times \left( \int_{V_{i-}} \boldsymbol{\sigma} \, dV \right)\cdot\mathbf{t}_{i-1},  
\end{equation}
where volume $V_{i-}$ is the half-bead joining contact point $c_{i-1}$
with bead center $\mathbf{r}_i$, whose pointing outward normal 
at $ \mathbf{r}_i$ is $\mathbf{t}_{i-1} = \mathbf{e}_{i-1,i}$.
The surface $S_{i-}$ enclosing   half-bead  $V_{i-}$ is composed of half-sphere $\mathcal{S}_{i-}$ and disk $\mathcal{D}_{i-}$, $S_{i-}=\mathcal{S}_{i-} \cup \mathcal{D}_{i-}$. Similarly, considering the moment balance equation (\ref{moment_balance_ll}) from bead center $\mathbf{r}_i$ to contact point $c_{i}$ leads to 
\begin{equation}
\label{tvecNi+1}
 \int_{2a(i+1/2)}^{2a(i+1)}\mathbf{t}\times\mathbf{n}^{is} \, ds = \mathbf{t}_{i}\times \int_{2a(i+1/2)}^{2a(i+1)}\mathbf{n}^{is} \, ds=\mathbf{t}_{i}\times \left( \int_{V_{i+}} \boldsymbol{\sigma} \, dV \right)\cdot\mathbf{t}_{i},  
\end{equation}
where volume $V_{i+}$ is the half-bead joining bead center $\mathbf{r}_i$ to contact point $c_i$, whose pointing outward normal at $ \mathbf{r}_i$ is $-\mathbf{t}_{i} = \mathbf{e}_{i+1,i}$.
The surface $S_{i+}$ enclosing   half-bead  $V_{i+}$ is composed of half-sphere $\mathcal{S}_{i+}$ and disk $\mathcal{D}_{i+}$, $S_{i+}=\mathcal{S}_{i+} \cup \mathcal{D}_{i+}$. Hence the integrated contribution of the second term of the moment balance equation (\ref{moment_balance_ll}) is the sum of the right-hand-side of (\ref{tvecNi}) and  (\ref{tvecNi+1}) which ought to be evaluated from  the volume integral of the total stress over $V_{i-} \cup V_{i+}$  inside  bead $i$. Since the stress tensor inside the beads is not known, it is possible to relate it to
the applied normal force at bead surface. Using divergence theorem on any volume $V$, enclosed by surface $S$, one finds 
\begin{equation}
\label{div}
\int_V  {\bf \sigma}_{\alpha\beta}  dV= \int_S  ({\bf \sigma}_{\alpha\gamma} \cdot  n_\gamma)   x_\beta  dS = \int_S  \boldsymbol{\sigma}  \cdot {\bf n}\otimes  {\bf x}  dS\equiv {\bf D}^{S},
\end{equation}
where $ {\bf n}$  is the normal pointing outward surface $S$, whilst introducing  the tensor ${\bf D}^{S}$ associated with the first moment contribution of the stress  at surface $S$. If the surface S is the surface enclosing the considered bead,  ${\bf D}^{S}$ is the usual stress tensor, associated with the hydrodynamic interactions between the fluid and the bead. When considering  hydrodynamic interactions, ${\bf D}^{S}$ is usually decomposed into a symmetric tensor called  stresslet and an  anti-symmetric one called couplet.
 Using  relation (\ref{div}) in (\ref{tvecNi}) as well as  (\ref{tvecNi+1}), one finds the following  four contributions
\begin{equation}
\label{tvecNi_tot}
 \int_{2a(i)}^{2a(i+1)}\mathbf{t}\times\mathbf{n}^{is} \, ds = \mathbf{t}_{i-1}\times  \left( {\bf D}^{\mathcal{S}_{i-}}+{\bf D}^{\mathcal{D}_{i-}}\right)  \cdot \mathbf{t}_{i-1}+\mathbf{t}_{i}\times \left( {\bf D}^{\mathcal{S}_{i+}}+{\bf D}^{\mathcal{D}_{i+}} \right) \cdot \mathbf{t}_{i},
\end{equation}
to the integration of the second term of (\ref{moment_balance_ll}). In the limit of bead radius tending to zero, then $\mathbf{t}_{i-1} \rightarrow \mathbf{t}_{i}$, so that
the outward normal vector to $\mathcal{D}_{i-}$, $\mathbf{t}_{i-1}$, tends to the opposite of the outward normal vector to $\mathcal{D}_{i+}$. Since $\mathcal{D}_{i-} \rightarrow \mathcal{D}_{i+}$, this implies in turn,  that
${\bf D}^{\mathcal{D}_{i-}} \rightarrow -{\bf D}^{\mathcal{D}_{i+}}$. Furthermore, since in the asymptotic limit of zero bead radius,  $\mathcal{S}_{i-} \cup  \mathcal{S}_{i+}  \rightarrow S$, one finds that

\begin{equation}
\label{tvecNi_tot}
 \int_{2a(i)}^{2a(i+1)}\mathbf{t}\times\mathbf{n}^{is} \, ds  \rightarrow \left( \mathbf{t}_{i-1}\times  ({\bf D}^{\mathcal{S}_{i-}}  \cdot \mathbf{t}_{i-1}) +\mathbf{t}_{i}\times  ({\bf D}^{\mathcal{S}_{i+}}  \cdot \mathbf{t}_{i}) \right)\rightarrow \mathbf{t}_{i}\times  ({\bf D}^{S}  \cdot \mathbf{t}_{i}),
\end{equation}
where $S$ is the bead surface here.
For now, we  concentrate on the contact forces contribution to  (\ref{tvecNi_tot}). Using the contact  surface force (\ref{boundary_condition}) it is then easy to compute the contact forces contribution to  (\ref{tvecNi_tot}),
\begin{equation}
\label{tvecNi_contact}
 \int_{2a(i)}^{2a(i+1)}\mathbf{t}\times\mathbf{n}^{is}_c \, ds \rightarrow  \left( a\mathbf{t}_{i-1}\times  {\bf f}_{c_{i-1}}  -a\mathbf{t}_{i}\times  {\bf f}_{c_i}  \right)=   \left( {\bf m}_{c_{i-1}}-{\bf m}_{c_i}  \right)
 \end{equation}

%The supplementary term to  (\ref{tvecNi_tot}) coming from hydrodynamic contributions related to stresslet and couplet  contributions is the following :
%\begin{equation}
%\label{tvecNi_couple_gamma}
%\mathbf{t}_{i}\times  ({\bf D}^{S}  \cdot \mathbf{t}_{i})=\epsilon_{\alpha \beta \gamma} t_{\beta} D^{S}_{\gamma \lambda} t_{\lambda} = \epsilon_{ikj} {\bf D}^{S}_{ik}=\epsilon_{ikj} \frac{1}{2}({\bf D}^{S}_{ik}- {\bf D}^{S}_{ki})={\gamma}^h_j\equiv\boldsymbol{\gamma}^h,
%\end{equation}
 %which only involves the  the anti-symmetric  couplet part of ${\bf D}^{S}$ is directly related to the hydrodynamic torque  $\boldsymbol{\gamma}^h$ associated with bead $i$ denoted $\boldsymbol{\gamma}^h_i$ in  (\ref{total_moment}).

%Nevertheless, as opposed to contact forces contributions which scale linearly with $a$, 
%those stresslet and rotlet  contributions
%scale as $a^3$. In the limit $a \rightarrow 0$,
%those terms are comparativelely much smaller and can
%therefore be safely neglected in the
%asymptotic limit, leading to a discrete form of
%(\ref{moment_balance_ll})   
%\begin{equation}
%\label{tvecNi_hydro}
%{\mathbf m}^{is}(2a(i+1))-{ \mathbf m}^{is}(2ai)+ \left( {\bf m}_{ci}-{\bf m}_{ci+1} \right )+\boldsymbol{\gamma}^h,
% \end{equation}
 Hence, result (\ref{tvecNi_contact}) is consistent
 with moment balance used in (\ref{total_moment}).

% Following citation commands can be used in the body text:


%\bibliographystyle{elsarticle-num}
% \bibliographystyle{elsarticle-harv}
% \bibliographystyle{elsarticle-num-names}
% \bibliographystyle{model1a-num-names}
% \bibliographystyle{model1b-num-names}
% \bibliographystyle{model1c-num-names}
 \bibliographystyle{model1-num-names}

\bibliography{JCP_Delmotte_Climent_Plouraboue_revised}

\end{document}